\documentclass[preprintnumbers,amsmath,amssymb]{revtex4}
\usepackage{graphicx}
\usepackage{dcolumn}% Align table columns on decimal point
\usepackage{bm}% bold math
\begin{document}
\def\oppropto{\mathop{\propto}} 
\def\opsimeq{\mathop{\simeq}}
\def\opoverderline{\mathop{\overline}}
\def\operarrow{\mathop{\longrightarrow}}
\def\opsim{\mathop{\sim}} 

\title{Distribution of pseudo-critical temperatures and lack of
self-averaging \\
in disordered Poland-Scheraga models with different loop exponents}
\author{ C\'ecile Monthus and Thomas Garel }
 \affiliation{Service de Physique Th\'{e}orique, CEA/DSM/SPhT\\
Unit\'e de recherche associ\'ee au CNRS\\
91191 Gif-sur-Yvette cedex, France}

\begin{abstract}

According to recent progresses in the finite size scaling theory of
disordered systems, thermodynamic observables are not self-averaging
at critical points when the disorder is relevant in the
Harris criterion sense. This lack of
self-averageness at criticality is directly related to the distribution
of pseudo-critical temperatures $T_c(i,L)$ over the ensemble of
samples $(i)$ of size $L$. In this paper, we apply this analysis to
disordered Poland-Scheraga models with different loop exponents $c$,
corresponding to marginal and relevant disorder. In all cases, we 
numerically obtain a Gaussian 
histogram of pseudo-critical temperatures $T_c(i,L)$ with mean
$T_c^{av}(L)$ and width $\Delta T_c(L)$. 
 For the marginal case $c=1.5$ corresponding to two-dimensional wetting, 
both the width $\Delta T_c(L)$ and the shift
$[T_c(\infty)-T_c^{av}(L)]$  decay as $L^{-1/2}$, so the exponent is
unchanged ($\nu_{random}=2=\nu_{pure}$) but disorder is relevant and
leads to non self-averaging at criticality.
For relevant disorder
$c=1.75$, the width $\Delta T_c(L)$ and the shift
$[T_c(\infty)-T_c^{av}(L)]$ decay with the same new exponent
$L^{-1/\nu_{random}}$ (where $\nu_{random} \sim 2.7 > 2 > \nu_{pure}$) and
there is again no self-averaging at criticality. Finally for the value
$c=2.15$, of interest in the context of DNA denaturation, the
transition is first-order in the pure case. In the presence of
disorder, the width $\Delta T_c(L) \sim L^{-1/2}$ 
dominates over the shift $[T_c(\infty)-T_c^{av}(L)] \sim L^{-1}$,
i.e. there are two correlation length exponents $\nu=2$ and $\tilde
\nu=1$ that govern respectively the averaged/typical loop distribution.

\bigskip

%PACS numbers: 87.14.Gg; 87.15.Cc; 82.39.Pj

\end{abstract}
\maketitle

\section{Introduction}

The stability of pure critical points with respect
to weak bond disorder is governed by the Harris criterion \cite{Harris} :
near a second order phase transition in dimension $d$,
the bond disorder is irrelevant if the specific heat exponent 
is negative $\alpha_{pure}=2-d \nu_{pure}<0$ or equivalently
if the correlation length exponent $\nu_P \equiv \nu_{pure} > 2/d$.
On the contrary if $\nu_{P} < 2/d$, disorder is relevant
and drives the system towards a random fixed point
characterized by a new correlation length
 exponent satisfying the general bound 
$\nu_R \equiv \nu_{random} \geq 2/d$ \cite{chayes}.
More recently, important progresses have been made in the understanding of
finite size properties of random 
critical points \cite{domany95,AH,Paz1,domany,AHW,Paz2}.
The main outcome of these
 studies can be summarized as follows.
To each disordered sample $(i)$ of size $L$, one should first associate
a pseudo-critical temperature $T_c(i,L)$, defined for instance
in magnetic systems as the temperature where the susceptibility is
maximum \cite{domany95,Paz1,domany,Paz2}. 
The disorder averaged pseudo-critical critical
temperature $T_c^{av}(L) \equiv \overline{T_c(i,L)}$
satisfies 
\begin{equation}
T_c^{av}(L)- T_c(\infty) \sim L^{-1/\nu_{R}}
\label{meantc}
\end{equation}
where $\nu_R$ is the correlation length
exponent. Eq. (\ref{meantc}) generalizes the analogous relation for
pure systems 
\begin{equation}
\label{puretc}
T_c^{pure}(L) - T_c(\infty) \sim L^{-1/\nu_{P}}
\end{equation}
The nature of the disordered critical point then depends on the
width $\Delta T_c(L)$ of the distribution of the pseudo-critical
temperatures $T_c(i,L)$
\begin{equation}
\Delta T_c(L) \equiv \sqrt{Var [T_c(i,L)]}
=\sqrt{\overline{T_c^2(i,L)}-\left(\overline{T_c(i,L)}\right)^2}
\end{equation}
When the disorder is irrelevant, the fluctuations
of these pseudo-critical temperatures obey
the scaling of a central limit theorem as in the Harris argument : 
\begin{equation}
\Delta T_c(L) \sim L^{-d/2} \ \ \hbox{ for  irrelevant  disorder }
\label{deltatcirrelevant}
\end{equation}
This behavior was first believed to hold in general \cite{domany95,Paz1}, 
but was later shown to be wrong in the case of random fixed points.
In this case, it was argued \cite{AH,domany} that
eq. (\ref{deltatcirrelevant}) should be replaced by
\begin{equation}
\Delta T_c(L) \sim L^{-1/\nu_{R}} \ \ \hbox{ for  random critical points }
\label{deltatcrelevant}
\end{equation}
i.e. the scaling is the same as the $L$-dependent shift of the averaged 
pseudo-critical temperature (Eq. \ref{meantc}).
The fact that these two temperature scales remain the same
is then an essential property of random fixed points
that leads to the lack of self-averaging for observables at criticality
\cite{AH,domany}. More precisely, for a given observable $X$,
it is convenient to define its normalized width as
\begin{equation}
\label{defratiodomany}
R_X(T,L) \equiv \frac{ \overline { X_i^2(T,L)} - ( \overline{X_i(T,L)})^2
}{ ( \overline{X_i(T,L)})^2 } 
\end{equation}
If $\xi(T)$ denotes the correlation length, the following behavior of
$R_X(T,L)$ is expected \cite{AH,domany}  

(i) off criticality, when $L \gg \xi(T)$, 
the system can be divided into nearly independent sub-samples
and this leads to `Strong Self-Averaging' 
\begin{equation}
R_X(T,L) \sim \frac{1}{ L^d} \ \ \hbox{ off  criticality  for 
 $L \gg \xi(T)$ } 
\end{equation}

(ii) in the critical region, when $L \ll \xi(T)$, 
the system cannot be divided anymore into nearly independent sub-samples.
In particular at $T_c(\infty)$ where $\xi=\infty$,
one can have either `Weak Self-Averaging' 
\begin{equation}
\label{weaksa}
R_X(T_c(\infty),L) \sim L^{ \frac{\alpha_P}{\nu_P}}  \ \ \hbox{ for  irrelevant
disorder ($\alpha_P <0$)  } 
\end{equation}
or `No Self-Averaging'
\begin{equation}
\label{nosa}
R_X(T_c(\infty),L) \sim Cst \ \ \hbox{ for   random  critical points  }
\end{equation}

In this paper, we study from this point of view 
disordered Poland-Scheraga models with different 
loop exponents $c$, corresponding to
either a pure second order transition with respectively
marginal/relevant disorder according to the Harris criterion,
or to a pure first-order transition.
In each case, we numerically 
compute the histogram of pseudo-critical temperatures
and study the self-averaging properties at criticality.
The paper is organized as follows. In Sec. \ref{PSmodels}, we recall
the definition of Poland-Scheraga (PS) models, the critical properties
of the pure transitions and the disorder relevance. In
Sec. \ref{observables}, we describe the observables that we
numerically compute for disordered PS models. We then present our
results for the different loop exponents considered, namely (i)
$c=1.5$ (marginal disorder) in Sec. \ref{c15} (ii) 
$c=1.75$ (relevant disorder) in Sec. \ref{c175} (iii) $c=2.15$ (case
of a first-order transition in the pure case) in Sec \ref{c215} .

\section{Poland-Scheraga models : disorder relevance at pure critical points}

\label{PSmodels}

\subsection{ Critical properties of pure Poland-Scheraga models} 

We first consider the adsorption (or wetting) transition of a polymer
chain onto an impenetrable substrate. This model corresponds to a
Poland-Scheraga model with loop exponent $c=1.5$. This wetting model
in dimension $1+1$ is defined as follows. The substrate is located at
$z=0$. The polymer chain has $L$ monomers, and the position
$z_{\alpha}$ of monomer $(\alpha)$ satisfies $z_{\alpha} \ge 0$, with
$z_1=z_L=0$ (bound-bound boundary conditions). The partition function
of the model reads 
\begin{equation}
\label{partit1}
Z=\sum_{(RW)} e^{-\beta H}
\end{equation}
where $H=\sum_{\alpha=1}^{L} \varepsilon_{\alpha}\
\delta_{z_{\alpha},0}$. In equation (\ref{partit1}), $\beta
=\frac{1}{k_B T}$ is the inverse temperature and the sum runs over all
walks (RW) such that $\vert z_{\alpha+1}-z_{\alpha} \vert =\pm
1$. This problem can be formulated \`a la
Poland-Scheraga \cite{Pol_Scher}
in terms of the loop statistics, where each loop of
length $l$ has a weight $\frac{2^l}{l^{3/2}}$.
We consider here generalized PS models where 
each loop of length $l$ has a weight $\frac{2^l}{l^c}$, with exponent $c>1$.

In the pure case $\varepsilon_{\alpha}=\varepsilon_0$,
the loop distribution at $T_c=T_c(\infty)$ is
\begin{equation}
\label{probaloop}
P^{pure}_{Tc} (l) \sim \frac{1}{l^c}
\end{equation}
For $c>2$, the averaged length $<l> = \int dl \ l P^{pure}_{Tc} (l) $
is finite, so that the number $n(T_c)$ of contacts with the substrate
is extensive ($n(T_c) \sim L$); the transition is 
therefore first order. For $1<c<2$, the averaged length $<l>$ diverges,
and the L\' evy sum of $n$ independent variables $l_i$ drawn from the
distribution (\ref{probaloop}) scales as $l_1+...+l_n \sim n^{1/(c-1)}$.
As a consequence at criticality, the number of contacts $n^{pure}_L(T_c)$
in a sample of length $L$ scales as
\begin{equation}
n_L^{pure}(T_c) \sim L^{c-1}
\end{equation}
and the transition is second order.
The contact density thus decays as
\begin{equation}
\theta_L^{pure}(T_c) = \frac{n_L^{pure}(T_c)}{L} \sim L^{c-2}
\label{thetatcpur}
\end{equation}
Its finite-size scaling form 
\begin{equation}
\label{fss}
\theta_L^{pure}(T)=  L^{c-2} Q \left[ (T-T_c)L^{\frac{1}{\nu_P}} \right]
\end{equation}
involves the crossover exponent $\phi_P=\frac{1}{\nu_P}=c-1$.

\subsection { Harris criterion on 2d order transitions $1<c<2$ }

The Harris criterion \cite{Harris} concerning the stability
of pure second order transitions with respect to the addition of disorder
relies on  the sign of the specific heat exponent 
\begin{equation}
\alpha_P=2-\nu_P = \frac{2c-3}{c-1}
\end{equation}
An equivalent way to decide whether disorder is relevant
consists in a simple power-counting analysis of the disorder 
perturbation exactly at $T_c$ :
the pure finite-size contact density $<\delta_{z_i,0}>_{pure} \sim L^{c-2}$
of Eq. (\ref{thetatcpur}) yields that the perturbation
due to the presence of a small disorder 
in the contact energies $\epsilon_i=\epsilon_0+\delta \epsilon_i$
scales as 
\begin{equation}
\label{powercounting}
 \sum_{i=1}^L \delta \epsilon_i <\delta_{z_i,0}>_{pure} \sim L^{1/2} \times L^{c-2}
= L^{c-\frac{3}{2}}
\end{equation}
Disorder is thus irrelevant for $1<c<\frac{3}{2}$
and relevant for $\frac{3}{2}<c<2$. The marginal case
$c=\frac{3}{2}$ has been debated for a long time
\cite{FLNO,Der_Hak_Van,Bhat_Muk,Ka_La,Cu_Hwa,Ta_Cha,wetting} 
and is of special interest since it corresponds to two-dimensional wetting
as explained above.
As a consequence, we have chosen to study in parallel the 
cases $c=1.5$ and $c=1.75$.

\subsection { Disorder relevance on 1st order transitions $c>2$ }

The effect of disorder on first-order transitions
has been discussed for a long time \cite{Im_Wo,Ai_We,Hu_Be,Cardy}.
In this respect, the most frequently studied system
\cite{Jac_Car,Ang_Igl} is the 
2D Potts model with $q>4$, for which the Aizenman-Wehr theorem
\cite{Ai_We} states that disorder changes the critical 
behavior from first order to second order.
The recent numerical study of this phenomenon
\cite{Ang_Igl,Mer_Ang_Igl} however shows that there are many subtleties :
in particular, the latent heat that 
 vanishes for continuous disorder, remains
finite for binary disorder. Disorder effects on Potts models
have also been studied in 3D \cite{Potts3d}.

Since all these studies consider
spin systems displaying coexisting domains in the pure case, their conclusions
cannot be directly applied to the PS-model for the following
polymeric reasons \cite{adn2005} :
in the PS model, there is no surface tension
between the localized and delocalized phases, and there exists 
an diverging correlation length $1/(T_c-T)$ in the pure case. 
So here, the simplest way to discuss the relevance of disorder
consists in the simple power-counting analysis of the disorder 
perturbation exactly at $T_c$ as in Eq. (\ref{powercounting}).
Here for $c>2$ where $<\delta_{z_i,0}>_{pure}$ is finite,
the perturbation
due to the presence of a small disorder 
in the contact energies $\epsilon_i=\epsilon_0+\delta \epsilon_i$ scales as
\begin{equation}
\label{powercounting2}
 \sum_{i=1}^L \delta \epsilon_i <\delta_{z_i,0}>_{pure} \sim L^{1/2} 
\end{equation} 
As a consequence, disorder is relevant.

In this paper, we consider the case $c=2.15$ where the pure transition
is first order, since this value is of interest for DNA denaturation
\cite{Barbara1,Carlon,Baiesi,Ka_Mu_Pe}. Furthermore, the effect of
disorder on this transition has been recently debated
\cite{Barbara2,adn2005,Gia_Ton}.   

\section{ Observables studied in disordered PS models }

\label{observables}

\subsection{ Numerical details }

As explained in details in our previous papers \cite{wetting,adn2005},
the PS models can be numerically studied via the recursion
relations satisfied by the partition function,
with a Fixman-Freire scheme \cite{Fix_Fre} for the entropy of loops.
All results presented in this paper have been obtained
for independent quenched random contact energies
($\varepsilon_{\alpha}$), distributed with a binary distribution.
For $1<c<2$, we have chosen
$\varepsilon=0$ or $\varepsilon=\varepsilon_{0}$ with probabilities 
$(1/2,1/2)$, with $\varepsilon_{0}(c=1.5)=-350K$ and
$\varepsilon_{0}(c=1.75)=-440K$, to obtain critical temperatures in
the same temperature range. 
For $c=2.15$ of interest in DNA denaturation, we have taken
the same values as in our previous study \cite{adn2005},
namely $\varepsilon=-355K$ or $\varepsilon=-390 K$
 with probabilities $(1/2,1/2)$.

The data we present correspond to various sizes $L$
with corresponding numbers $n_s(L)$ of independent samples.
Unless otherwise stated, we have considered the 
following sizes going from $L=2 \cdot 10^3$ to $L=2048 \cdot 10^3 $, 
with respectively $n_s(L=2 \cdot 10^3)=3.84 \cdot 10^6 $ to $n_s(L=2048 \cdot
10^3)= 15 \cdot 10^3$. More precisely, we consider
\begin{eqnarray}
\label{sizesandnseed}
 \frac{L}{1000} && = 2, 4 , 8, 16, 32, 64, 128 , 256, 512, 1024, 2048   \\
  \frac{n_s(L)}{1000} && = 3840, 1920 , 960 , 480, 240, 120, 60 , 30, 15, 
30 ,15 
\end{eqnarray}

\subsection{Definition of a sample-dependent pseudo-critical temperature}
\label{deftcil}

In the magnetic systems studied in \cite{domany95,domany}, the pseudo-critical
temperature $T_c(i,L)$ of the sample $i$ was identified to the maximum
of the susceptibility. In the PS model, one can not follow the same
path and we have tried two different definitions :

{\it Definition of a pseudo-critical temperature
from the free-energy}

In the pure PS model with bound-bound boundary conditions, 
the behavior of the partition function as a function of temperature reads
 \begin{eqnarray}
Z_L^{pure}&&( T<T_c) \opsimeq_{L \gg 1/(T-T_c)^{\nu_P}}
  (T_c-T)^{\nu_P-1} 2^L e^{ (T_c-T)^{\nu_P} L } \nonumber \\
 Z_L^{pure}&&(T_c) \simeq \frac{ 2^L }{ L^{2-c} }  
\label{znpur}  \\
Z_L^{pure}&&(T>T_c) \opsimeq_{L \gg 1/(T-T_c)^{\nu_P} }  \nonumber
 \frac{ 2^L }{ (T-T_c)^2 L^{c} } 
\end{eqnarray}
with $\nu_P=1/(c-1)$.
A finite-size pseudo-critical temperature $T_c^{pure(f)}(L)$
can then be defined as the temperature where the free-energy
reaches the extensive delocalized value $F_{deloc}=-T L \ln 2$, i.e.
$T_c^{pure(f)}(L)$ is the solution of the equation
 \begin{eqnarray}
F_L^{pure}(L,T)+T L \ln 2=0
\end{eqnarray}
This definition from the free-energy is very natural,
but has the drawback of introducing a logarithmic factor
\begin{equation}
\label{lnfre}
T_c^{pure(f)}(L) - T_c(\infty) \sim \left( \frac{\ln L}{L} \right)^{1/\nu_{P}}
\end{equation}
with respect to the purely algebraic factor usually expected (Eq. \ref{puretc}).
This logarithmic factor comes the finite-size free-energy
value exactly at criticality $F_L^{pure}(L,T_c)=-T_c L \ln 2 +(2-c) T_c \ln L$
(Eq. \ref{znpur}). In the disordered case, we may similarly define a
sample-dependent pseudo-critical temperature $T_c^{(f)}(i,L)$ as the
solution of the equation
 \begin{eqnarray}
F_L^{(i)}(L,T)+T L \ln 2=0
\end{eqnarray}
but logarithmic corrections are to be expected 
at least in the shift (Eq. \ref{meantc}). Since these logarithmic
factors may alter the numerical measures of critical exponents in the
disordered cases, we have also  considered another definition of
$T_c(i,L)$.    

{\it Definition of a pseudo-critical temperature
from a sample-replication procedure}

In the pure PS model with $1<c<2$, another finite-size critical temperature
$T_c^{pure (\theta)}(L)$ may be defined as the temperature 
where the appropriately rescaled contact densities cross (Eq. \ref{thetatcpur})
 \begin{eqnarray}
L^{2-c} \theta^{pure}_L \big[ T_c^{pure (\theta)}(L) \big]=
(2L)^{2-c} \theta^{pure}_{2L}\big[T_c^{pure (\theta)}(2L)\big]
\end{eqnarray}
This temperature $T_c^{pure (\theta)}(L)$
 follows the usual algebraic behavior (Eq. \ref{puretc}),
in contrast with the other definition $T_c^{pure(f)}(L)$
that introduces logarithmic corrections (\ref{lnfre}).

This definition can be extended to the disordered case
via the following strategy first introduced in
\cite{adn2005} for the case $c=2.15$ (with bound-unbound boundary
conditions ). In short, one considers a sample
$(i)$ of length $L$, the sample $(2i)$ of length $2L$, obtained by
gluing together two copies of sample $i$,
as well as the sample $(4i)$ obtained by
gluing together four copies of sample $i$ : 
the three contact densities of $(i)$, $(2i)$ and $(4i)$
cross at a temperature (see Fig. 9b  of \cite{adn2005}) that can be interpreted
as the pseudo-critical temperature $T_c(i,L)$ of the disordered sample $(i)$
of length $L$.

Here we generalize this procedure to the case $1<c<2$.
The basic idea is that in disordered Poland-Scheraga models,
the loop distribution has exponent $c$ at criticality
in the thermodynamic limit as in the pure case (\ref{probaloop}).
We have found numerically this property in our previous studies
on the wetting transition with $c=1.5$ \cite{wetting}
and on the DNA denaturation with $c=2.15$ \cite{adn2005}.
 This 
suggests that the density of contacts scales as $L^{2-c}$
at criticality as in (\ref{thetatcpur}).

\begin{figure}[htbp]
%\begin{figure}
\includegraphics[height=6cm]{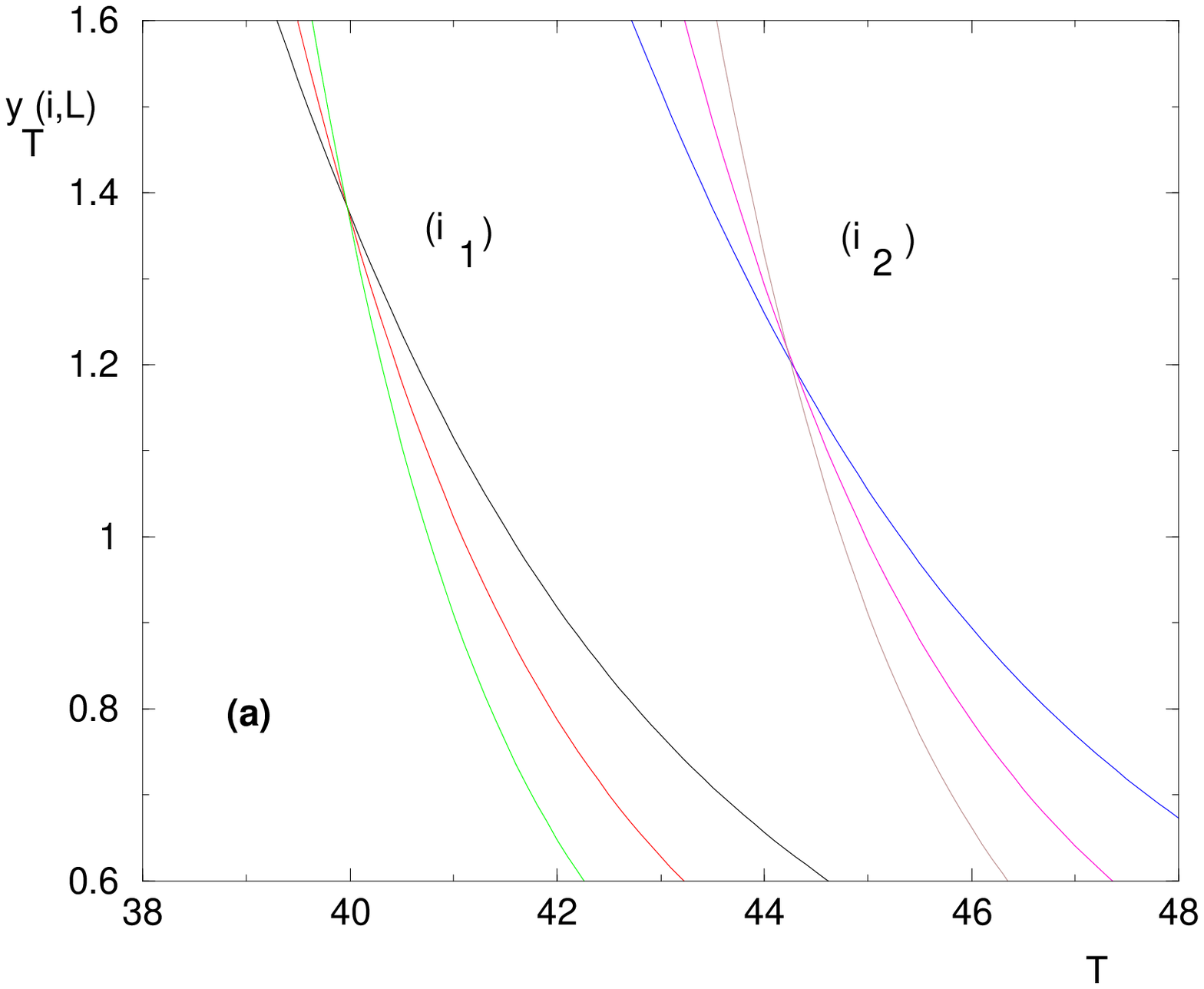}
\hspace{1cm}
\includegraphics[height=6cm]{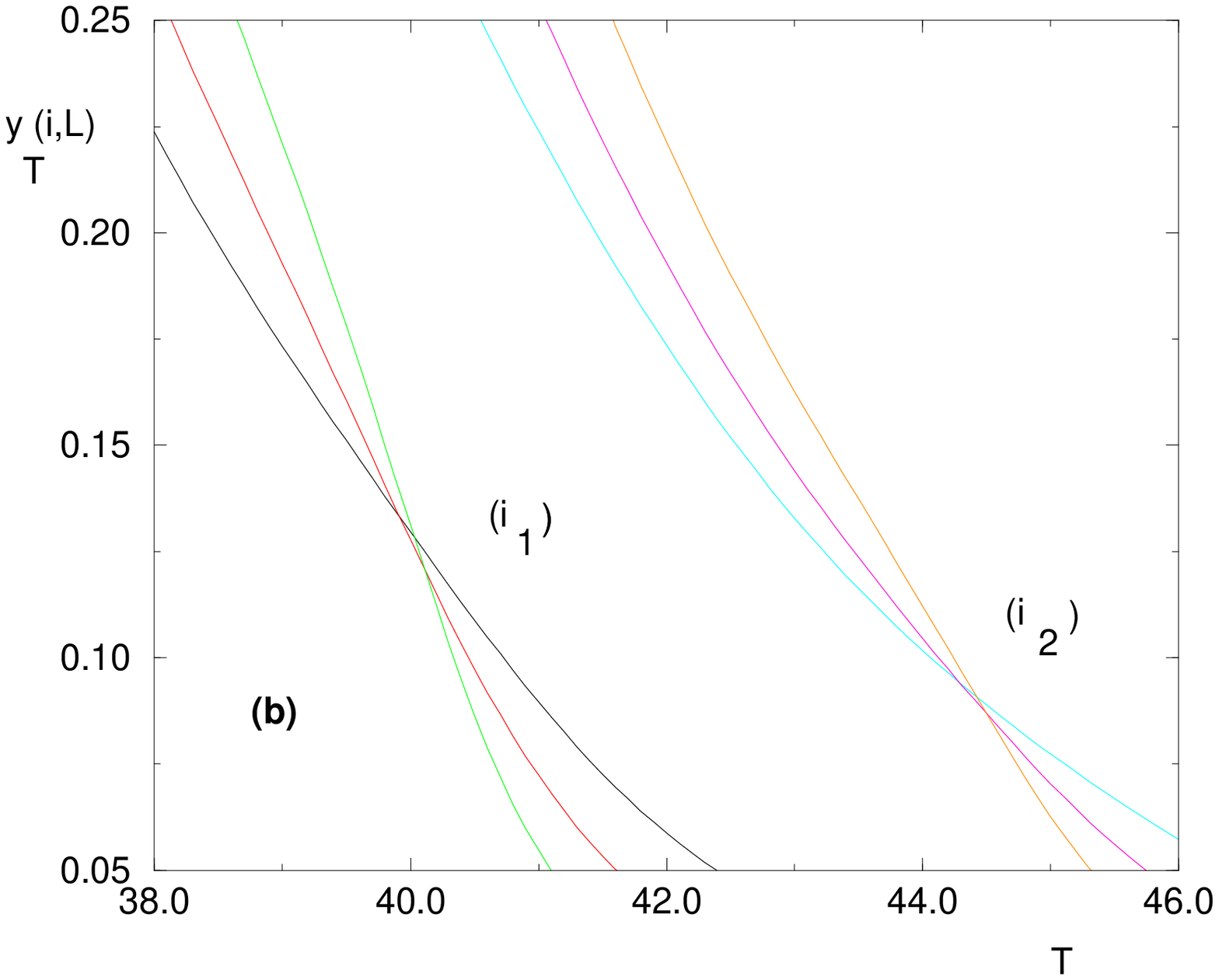}
\caption{ Illustration of the crossing procedure to obtain the pseudo-critical
temperatures $T_c(i,L)$ for two samples $i=i_1,i_2$
(a) for the case c=1.5 with L=50000 (b) for the case $c=1.75$ with $L=100000$.}
\label{crossing124c15}
\end{figure}

For $1<c<2$, we have thus tried to define a sample-dependent
pseudo-critical temperature $T_c(i,L)$ as follows.
For each sample
$(i)$ of length $L$, we construct the sample $(2i)$ of length $2L$
by gluing together two copies of sample $i$,
as well as the sample $(4i)$ by
gluing together four copies of sample $i$.
We then plot the rescaled contact densities 
\begin{equation}
\label{ycrossing}
y_T(i,L)=\theta_T(i,L) \ L^{2-c}
\end{equation}
with its two analogs $y_T(2i,2L)= \theta_T(2i,2L) \ (2L)^{2-c}$
and $y_T(4i,4L)= \theta_T(4i,4L) \ (4L)^{2-c}$.
Typical results obtained respectively for the cases $c=1.5$ and $c=1.75$
with two disordered samples $i=i_1,i_2$ 
are shown on Fig. \ref{crossing124c15} :
for each sample $(i)$, the three rescaled contact densities 
$y_T(i,L)$, $y_T(2i,2L)$ and $y_T(4i,4L)$ cross
at a temperature that we defined as 
the pseudo-critical temperature $T_c^{(\theta)}(i,L)$
of the disordered sample $i$ of length $L$. This crossing
of three curves validates the 
present replication procedure to define a proper pseudo-critical
temperature.

{\it Discussion}

Each definition of $T_c(i,L)$ has its own advantage and drawback. The
definition $T_c^{(f)}(i,L)$, that uses the free-energy, is probably
the most natural one, but introduces logarithmic corrections, already
in the pure PS model. The definition $T_c^{(\theta)}(i,L)$, that uses
the sample-replication procedure, may appear more artificial in the
disordered case, but it does not introduce logarithmic corrections.
We have checked numerically that both definitions
actually yield very similar results for critical exponents and scaling
distributions, even though they give different values for a given sample.
This shows that the conclusions that can be obtained 
do not depend on the precise definition of the pseudo-critical temperature.

\subsection{Distribution of pseudo-critical temperatures}

In all cases ($c=1.5$, $1.75$ and $2.15$),
and for both definitions of the pseudo-critical temperature
(either from the free-energy or from the sample-replication procedure),
we numerically obtain that  
the distributions of pseudo-critical temperatures
follows the scaling form
\begin{equation}
P_L(T_c(i,L)) \simeq  \frac{1}{ \Delta T_c(L)} \  g \left( x= \frac{
T_c(i,L) -T_c^{av}(L)}{ \Delta T_c(L) }  \right) 
\label{rescalinghistotc}
\end{equation}
where the scaling distribution $g(x)$ is simply Gaussian
\begin{equation}
g(x)=  \frac{1}{ \sqrt{2 \pi} } e^{- x^2/2}
\end{equation}
This means that the only important properties
of the pseudo-critical temperatures distribution
are the behaviors of its average $T_c^{av}(L)$ 
and width $\Delta T_c(L)$  as $L$ varies.

For the average $T_c^{av}(L)$, we have fitted our data
with the power-law (Eq.\ref{meantc}) for 
the temperature $T_c^{(\theta)}(i,L)$,
and with the generalized form involving logarithm
 (see Eq. \ref{lnfre} for the pure case)
for the temperature $T_c^{(f)}(i,L)$
\begin{eqnarray}
T_c^{av(\theta)}(L)- T_c(\infty)
 &&\simeq -A_{\theta}  \left(\frac{1}{L} \right)^{1/\nu_{R}} \\
T_c^{av (f)}(L)- T_c(\infty)  &&\simeq 
-A_{f}  \left(\frac{\ln L}{L} \right)^{1/\nu_{R}}
\end{eqnarray}
The two definitions of $T_c(i,L)$ then
yield compatible estimates for the exponent $\nu_R$
and the critical temperature $T_c(\infty)$.

For the variance, we have fitted our data with 
the power-law (Eq. \ref{deltatcrelevant})
for both definitions of pseudo-critical temperatures,
and the two definitions of $T_c(i,L)$ then
yield compatible estimates for the corresponding exponent.

\subsection{ Study of self-averageness in the critical region}

As already mentioned in the Introduction,
the issue of self-averageness at random critical points
is directly related to the scale of the width
of the pseudo-critical temperatures \cite{AH,domany} :
if the width $\Delta T_c(L)$ is of the same order of
the shift $T_c^{av}(L)-T_c(\infty)$, there is no self-averaging
at $T_c(\infty)$.

We have thus measured for PS-models
 the $L$ dependence of the ratio $R_{\theta}(T,L)$
defined in Eq. (\ref{defratiodomany}) for the contact density
for temperature $T$ in the critical region
\begin{equation}
\label{ratiotheta}
R_{\theta}(T,L) \equiv
\frac{Var[\theta_i(T,L)]}{(\overline{\theta_i(T,L)})^2 } 
\end{equation}

We have also studied the issue of self-averageness at $T_c(i,L)$ 
via the measure of the ratio
\begin{equation}
\label{defrcl}
{\cal R}_c(L) \equiv \frac{ Var [\theta_i(T_c(i,L),L)] }
{ (\overline {\theta_i(T_c(i,L),L)})^2}
\end{equation}

We now make contact with
the finite size scaling Ansatz of
\cite{domany95,domany} for an observable $X$ 
\begin{equation}
X_L^{(i)}(T) = L^{\rho} Q_i \left( (T-T_c(i,L)) L^{1/\nu_{R}} \right)
\label{fssdomany}
\end{equation}
where sample dependence arises through both $T_c(i,L)$ and the scaling
function $Q_i$. The ratio $R_{\theta}(T,L)$ mostly tests the relevance of
the variance $\Delta T_c(L)$, whereas
the ratio ${\cal R}_c(L) $ of Eq. (\ref{defrcl})
directly tests the sample dependence of $Q_i(0)$ for the observable
$X=\theta$. We have measured the ratio ${\cal R}_c(L) $ for both
definitions of $T_c(i,L)$. Although numerically different, their
qualitative behaviour is the same: whenever disorder is relevant, the
ratio ${\cal R}_c(L) $ grows with $L$.

We now present our results for the different loop exponents $c$.

\section{ The case of marginal disorder ($c=1.5$)}
\label{c15}

The marginal 
case $c=1.5$ corresponds to two-dimensional wetting
and has been the subject of a long-standing debate
\cite{FLNO,Der_Hak_Van,Bhat_Muk,Ka_La,Cu_Hwa,Ta_Cha,wetting}.
On the analytical side, efforts have focused on the
small disorder limit : Ref \cite{FLNO} finds a marginally
irrelevant disorder where
 the quenched critical properties are the same as in the pure case,
 up to subdominant logarithmic corrections. Other
studies have concluded that that the disorder is marginally relevant
\cite{Der_Hak_Van,Bhat_Muk,Ka_La}. On
the numerical side, the same debate on the disorder 
relevance took place. The numerical studies of Ref. \cite{FLNO}
and Ref. \cite{Cu_Hwa} have concluded that the
critical behavior was indistinguishable from the pure transition. On
the other hand, the numerical study of \cite{Der_Hak_Van}
 pointed towards a negative specific heat exponent
($\alpha<0$), and finally Ref. \cite{Ta_Cha}
has been interpreted as an essential singularity in the specific heat,
that formally corresponds to an exponent $\alpha=-\infty $. 

It is thus very interesting to study the marginal case $c=1.5$
 from the point of view
of the histogram of pseudo-critical temperatures and of
self-averaging properties to clarify the situation.

\subsection{Distribution of pseudo-critical temperatures}

\begin{figure}[htbp]
%\begin{figure}
\includegraphics[height=6cm]{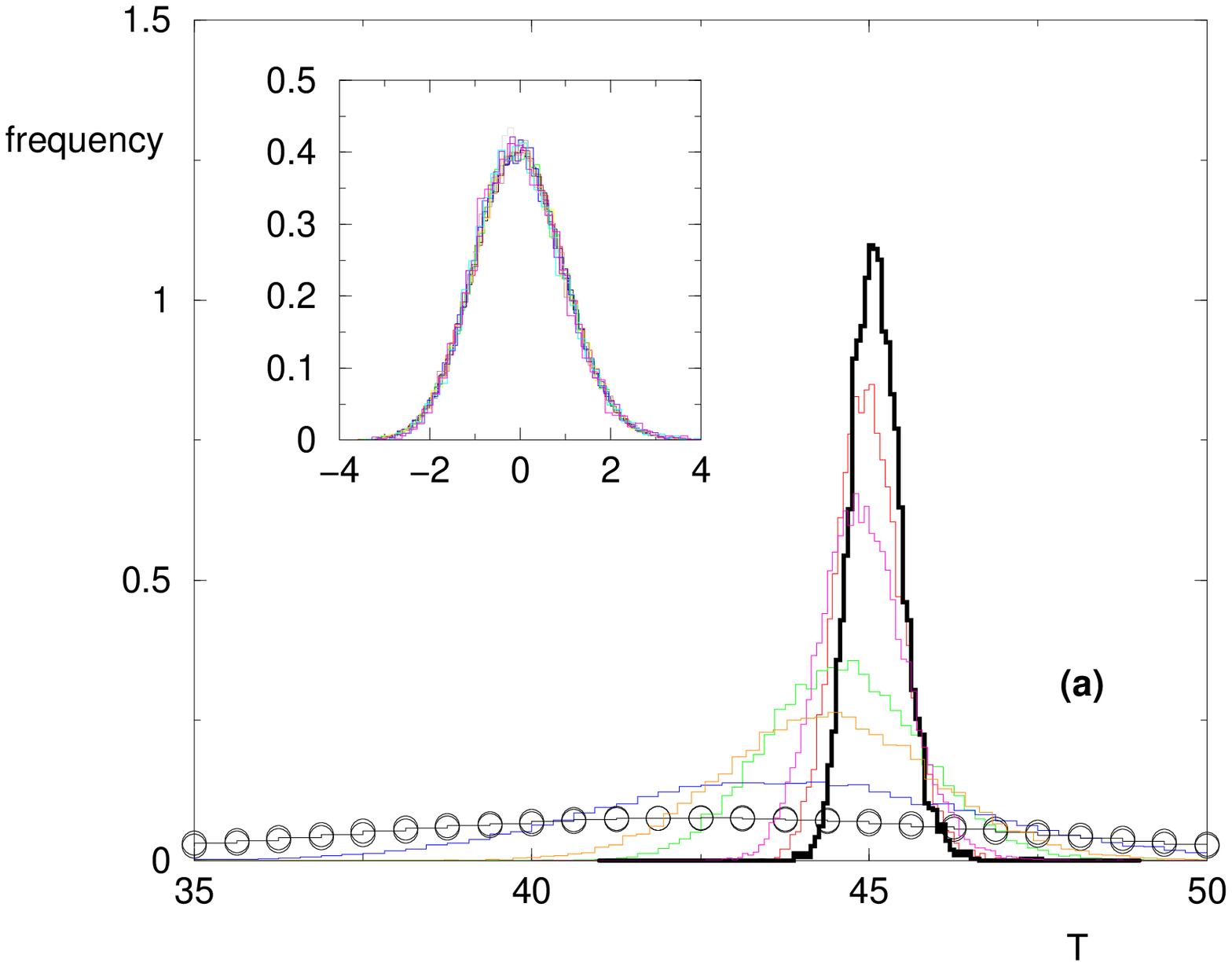}
\hspace{1cm}
\includegraphics[height=6cm]{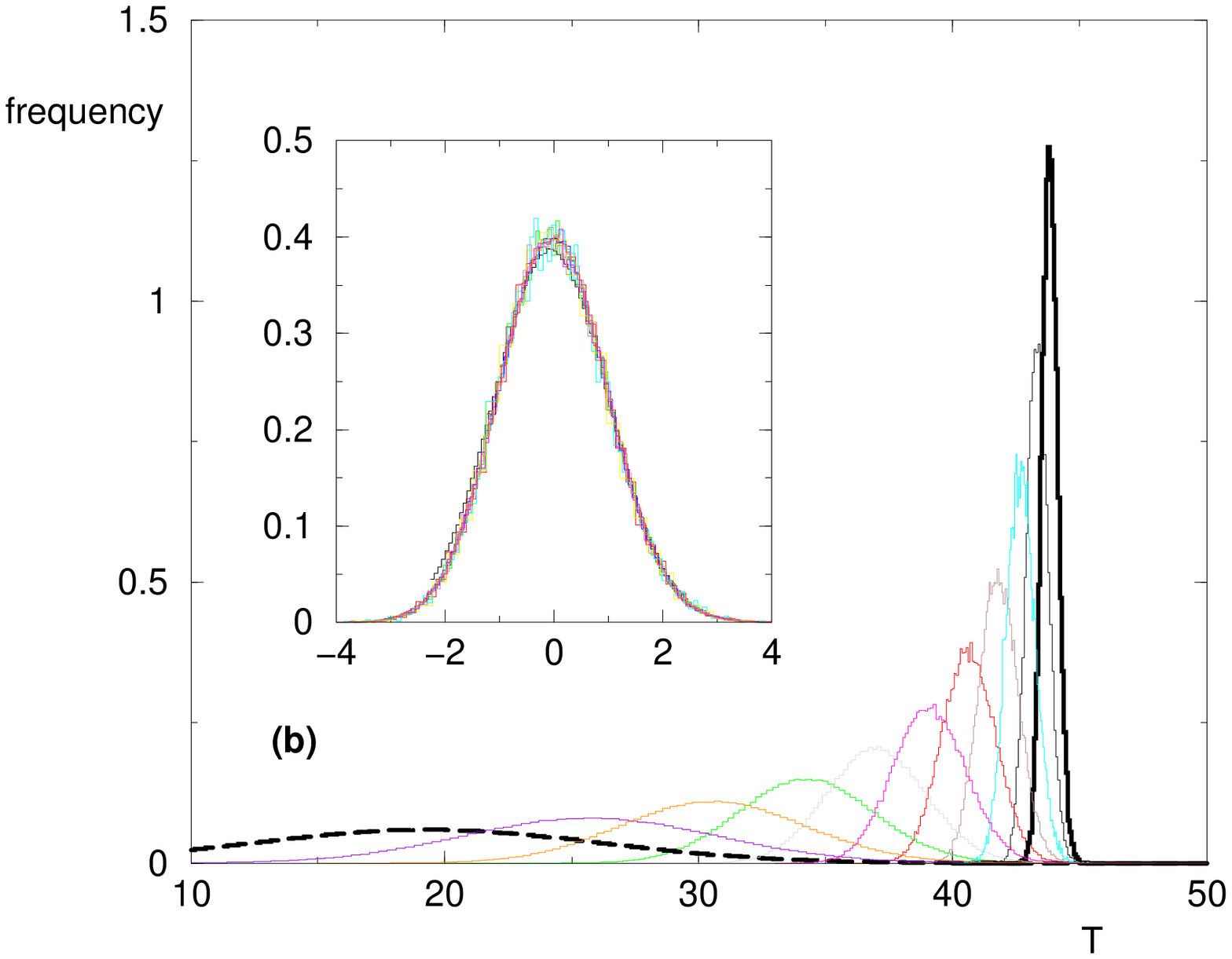}
\caption{ Case c=1.5 :
 Distribution of pseudo-critical temperatures $T_c(i,L)$,
together with the rescaling of Eq. (\ref{rescalinghistotc}) in Inset.
(a) for definition $T_c^{(\theta)}(i,L)$ with samples size
from $L/1000=4 (\bigcirc)$ 
% 16, 64,128, 512, 1024,
to $2048 $ (thick line).
(b) for definition $T_c^{(f)}(i,L)$ with samples size
from $L/1000= 2$ (dashed line) to $2048$ (thick line) } 
\label{figc15histotc}
\end{figure}

The histograms of both definitions of the pseudo critical temperatures
$T_c(i,L)$ are shown on Figure \ref{figc15histotc}(a) and (b). 
As mentioned previously, both rescaled distributions shown in the
Insets of Figure \ref{figc15histotc} are Gaussian
(Eq. \ref{rescalinghistotc}).

The averages $T_c^{av}(L)$ behave as
\begin{eqnarray}
T_c^{av(\theta)}(L) && \simeq 45.15- 132 \left(\frac{1}{
L}\right)^{0.47} \nonumber\\ 
T_c^{av(f)}(L)  && \simeq 45.25- 335 \left(\frac{\ln L}{L}\right)^{0.46}
\label{shiftc15}
\end{eqnarray}
These values of $T_c(\infty) \sim 45.2$ agree
with a different determination via the loop statistics in \cite{wetting}.
The mesaured exponent $\nu_R$ is very close to the pure exponent
$\nu_P=2$. However, this does not mean that the disorder is irrelevant,
since power-law fits of the variances yield
\begin{eqnarray}
\Delta T_c^{(\theta)}(L) && \simeq  206 \left(\frac{1}{
L}\right)^{0.44} \nonumber\\
\Delta T_c^{(f)}(L) && \simeq  188 \left(\frac{1}{
L}\right)^{0.44}
\label{widthc15}
\end{eqnarray}
Note that disorder being marginal, extra logarithmic factors could be
present in the shift and in the variance. Theoretical predictions on
this point are however unavailable, and we have not tried to include
these extra factors in the above fits. 

Our conclusion is that the shift and the width have very close
exponents, pointing towards a random critical point
(Eq. \ref{deltatcrelevant}) with non self-averaging properties that we
now consider.

\subsection{ Non-self-averaging properties }

\begin{figure}[htbp]
%\begin{figure}
\includegraphics[height=6cm]{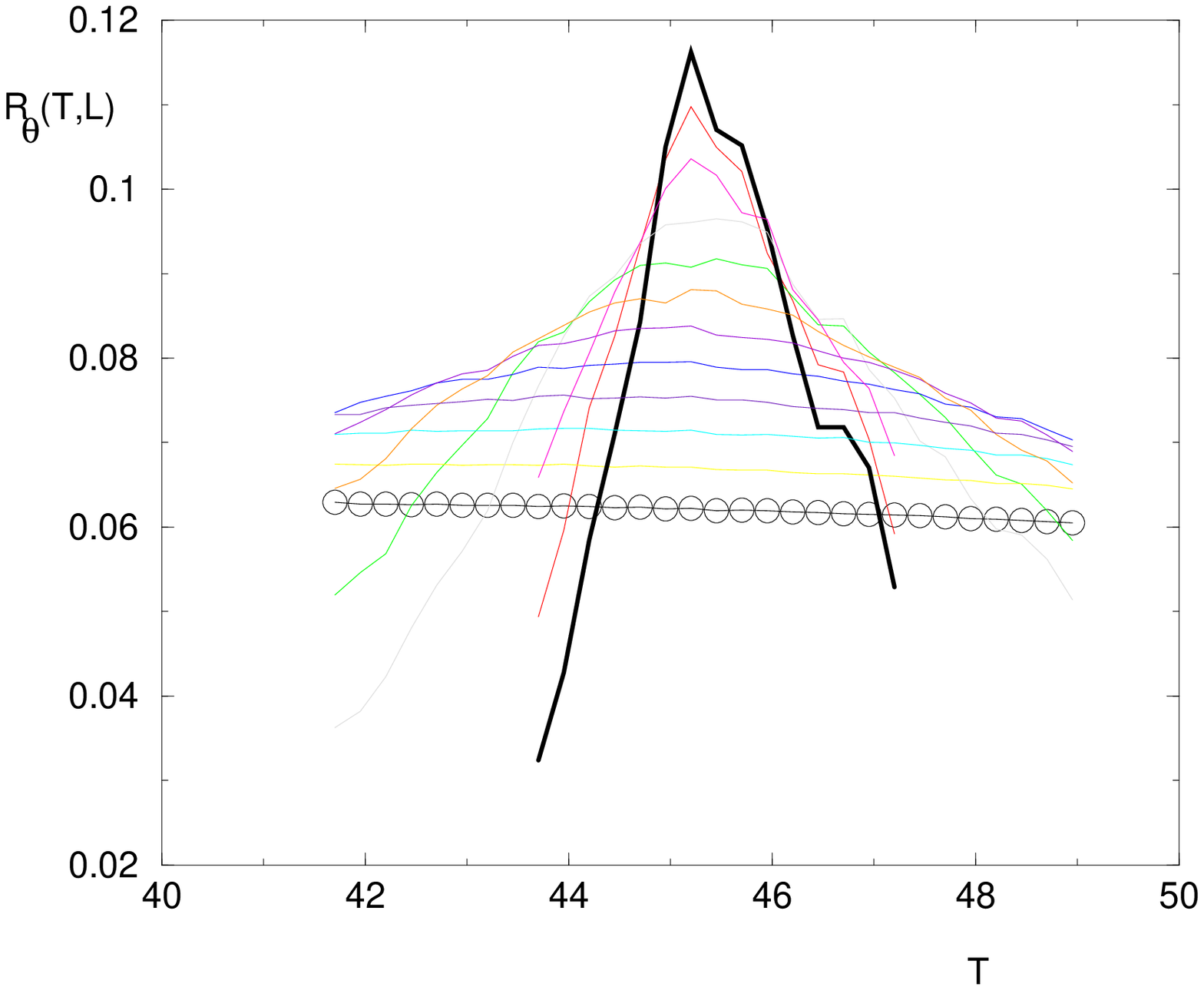}
\hspace{1cm}
\includegraphics[height=6cm]{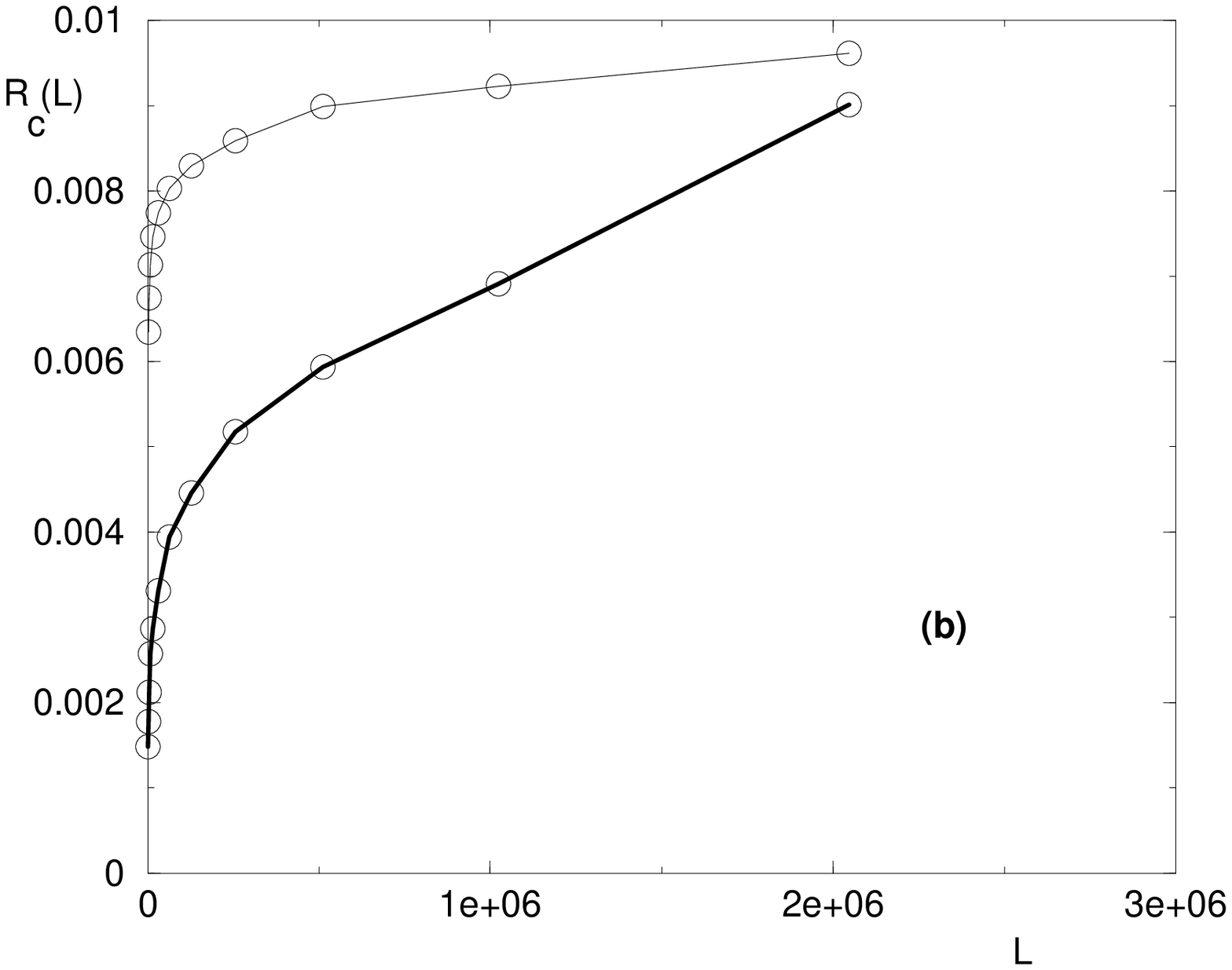}
\caption{ Case c=1.5 :
(a) Ratio $R_{\theta}(T,L)$ for sizes $L/1000=1(\bigcirc)$
% ,2, 4, 8, 16, 32, 64,128, 256,512, 1024, 
to $2048 $ (thick line)
(b) The ratios ${\cal R}_c(L)$ of (Eq. \ref{defrcl}), corresponding to
definitions $T_c^{(\theta)}(i,L)$ (thick line) and 
$T_c^{(f)}(i,L)$ (thin line), both grow with $L$.} 
\label{figc15ratio}
\end{figure}

We show on Figure \ref{figc15ratio} (a) the behavior of
the ratio $R_{\theta}(T,L)$ defined in (Eq. \ref{ratiotheta}). 
Near the critical temperature $T_c(\infty) \sim 45.2$ obtained in
Eq.\ref{shiftc15}, this ratio clearly displays a lack of self
averaging, as expected at random critical points (Eq. \ref{nosa}).

We show on Figure \ref{figc15ratio} (b) the ratios ${\cal R}_c(L)$
of Eq. (\ref{defrcl}) corresponding to the two definitions of
$T_c(i,L)$. They both grow with $L$.

\subsection{ Conclusion on the nature of the transition for $c=1.5$}

Our numerical results (\ref{shiftc15},\ref{widthc15}) indicate that in
the presence of disorder, the shift- and width- exponents 
are close to each other and to the pure case exponent
$\frac{1}{\nu_{pure}}=0.5$. We thus conclude
\begin{equation}
\nu_{random}=2=\nu_{pure}
\end{equation}
The disorder is nevertheless relevant in the sense that
there is a lack of self-averaging properties 
at criticality as in Eq. (\ref{nosa}),
because both the width $\Delta T_c(L)$ and the shift
$[T_c(\infty)-T_c^{av}(L)]$ decay with the same exponent
 \begin{equation}
\Delta T_c(L) \sim [T_c(\infty)-T_c^{av}(L)] \sim L^{-1/2}
\end{equation}

\section{ The case of relevant disorder ($c=1.75$)}
\label{c175}

\subsection{Distribution of pseudo-critical temperatures}

\begin{figure}[htbp]
%\begin{figure}
\includegraphics[height=6cm]{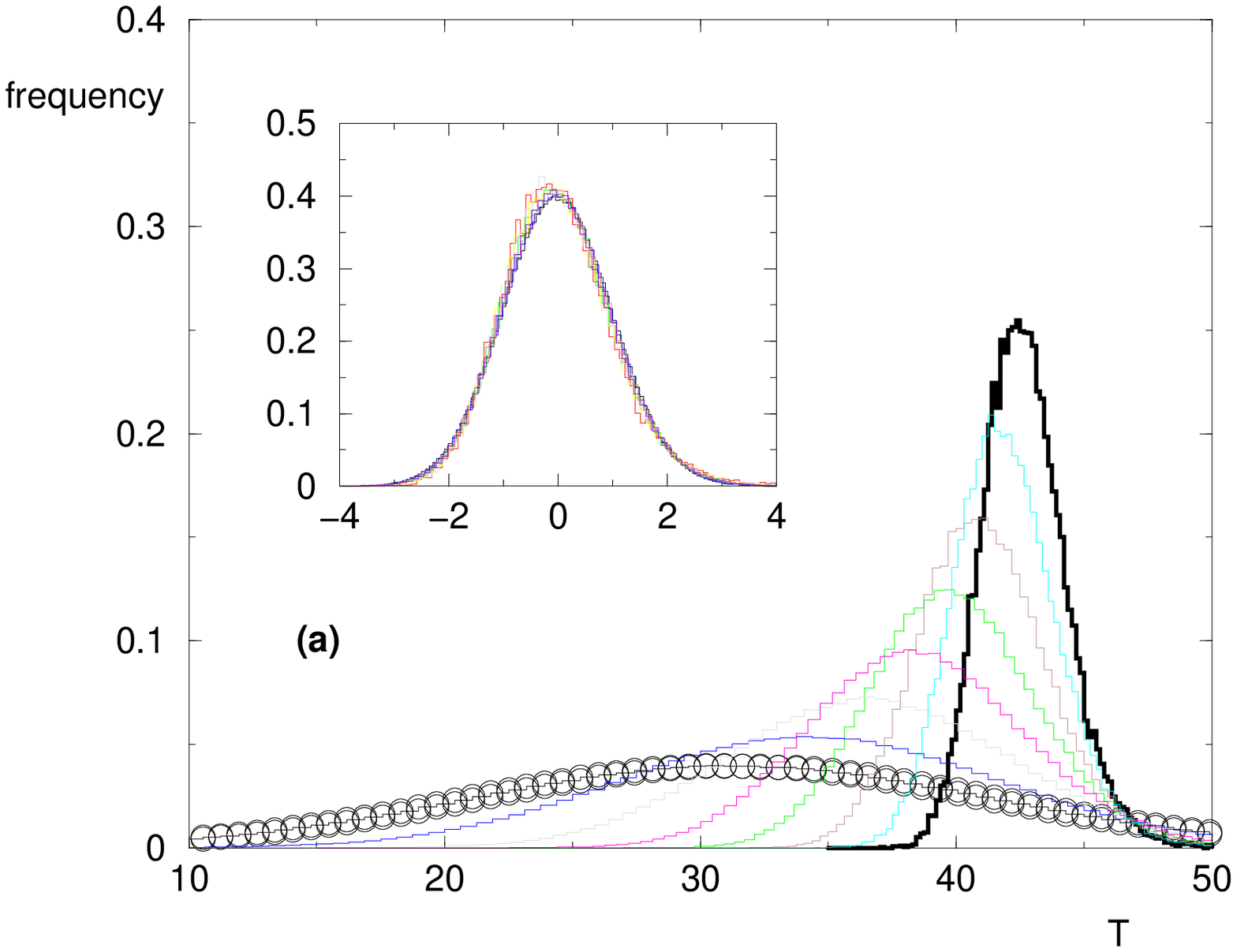}
\hspace{1cm}
\includegraphics[height=6cm]{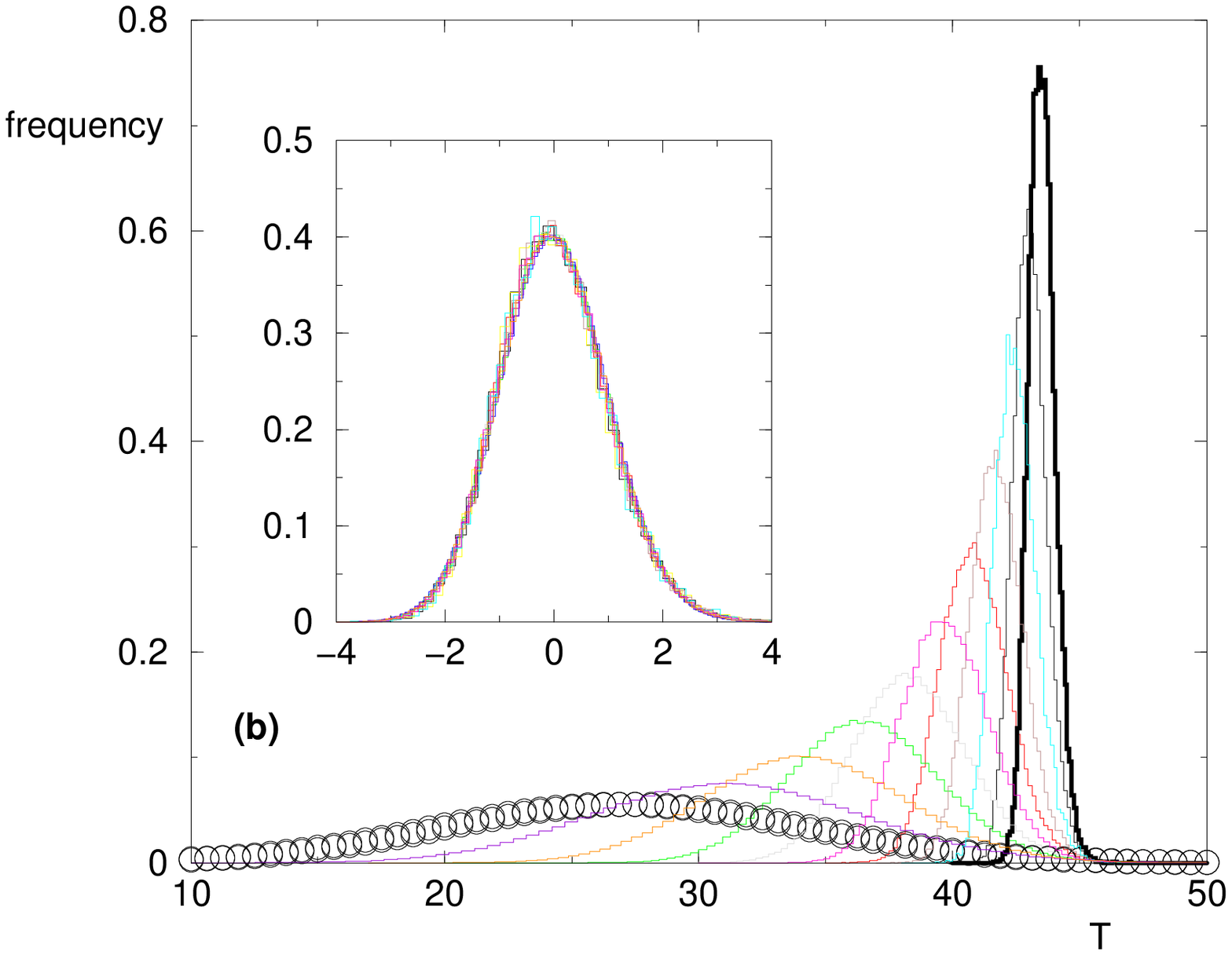}
\caption{ Case c=1.75 :
 Distribution of pseudo-critical temperatures $T_c(i,L)$,
together with the rescaling of Eq. (\ref{rescalinghistotc}) in Inset.
(a) for definition $T_c^{(\theta)}(i,L)$ with samples size
from $L/1000=1 (\bigcirc)$
% ,2, 4, 8, 16, 32, 64,
to $128$ (thick line).
(b) for definition $T_c^{(f)}(i,L)$ with samples size
from $L/1000=2 (\bigcirc)$ to $2048$ (thick line) } 
\label{figc175histotc}
\end{figure}

The histograms of both definitions of the pseudo critical temperatures
$T_c(i,L)$ are shown on Figure \ref{figc175histotc}(a) and (b). 
Both rescaled distributions shown in the
Insets of Figure \ref{figc175histotc} are Gaussian
(Eq. \ref{rescalinghistotc}).

The respective averaged $T_c^{av}(L)$ behave as
\begin{eqnarray}
\label{shiftc175}
T_c^{av(\theta)}(L) && \simeq 45.2- 155 \left(\frac{1}{
L}\right)^{0.35} \nonumber\\
T_c^{av(f)}(L)  && \simeq 45.4- 140 \left(\frac{\ln
L}{L}\right)^{0.37}
\end{eqnarray}

This shift with the size $L$ 
shows (Eq. 1\ref{meantc}) that the exponent $1/\nu_R=0.35$ 
is very different from the pure exponent
$1/\nu_P=c-1=0.75$. As expected,
the random exponent satisfies the general bound $\nu_R \geq 2 $ \cite{chayes}.
For both definitions the widths decay with the same exponent, namely
\begin{eqnarray}
\Delta T_c^{(\theta)}(L) && \simeq  134 \left(\frac{1}{
L}\right)^{0.38} \nonumber\\
\Delta T_c^{(f)}(L) && \simeq  117 \left(\frac{1}{
L}\right)^{0.38}
\end{eqnarray}
in agreement with the prediction of Eq. (\ref{deltatcrelevant})
\begin{equation}
\Delta T_c(L) \sim  T_c^{av}(L) - T_c(\infty)
\end{equation}
for random critical points.

\subsection{ Non-self-averaging properties}

\begin{figure}[htbp]
%\begin{figure}
\includegraphics[height=6cm]{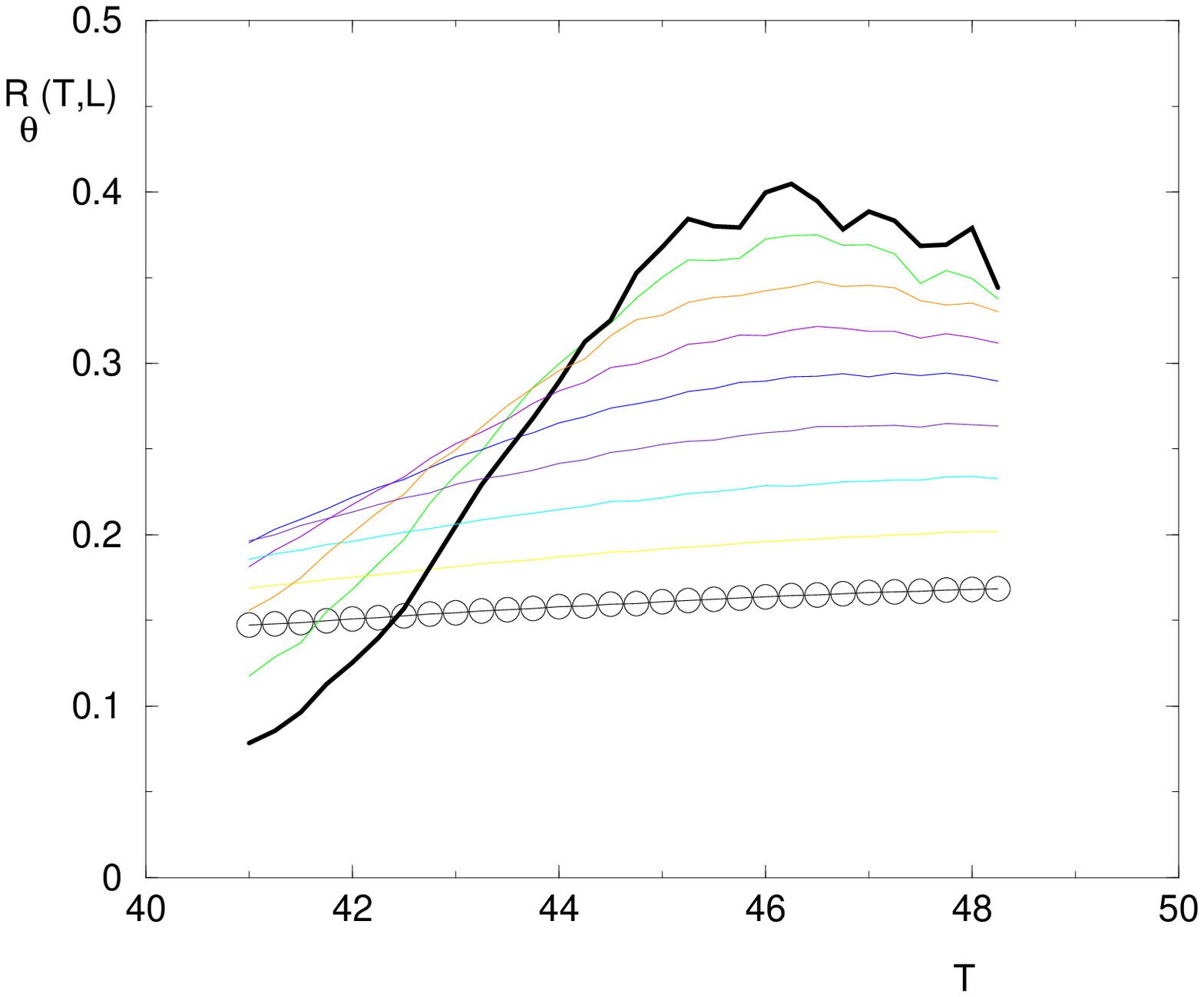}
\hspace{1cm}
\includegraphics[height=6cm]{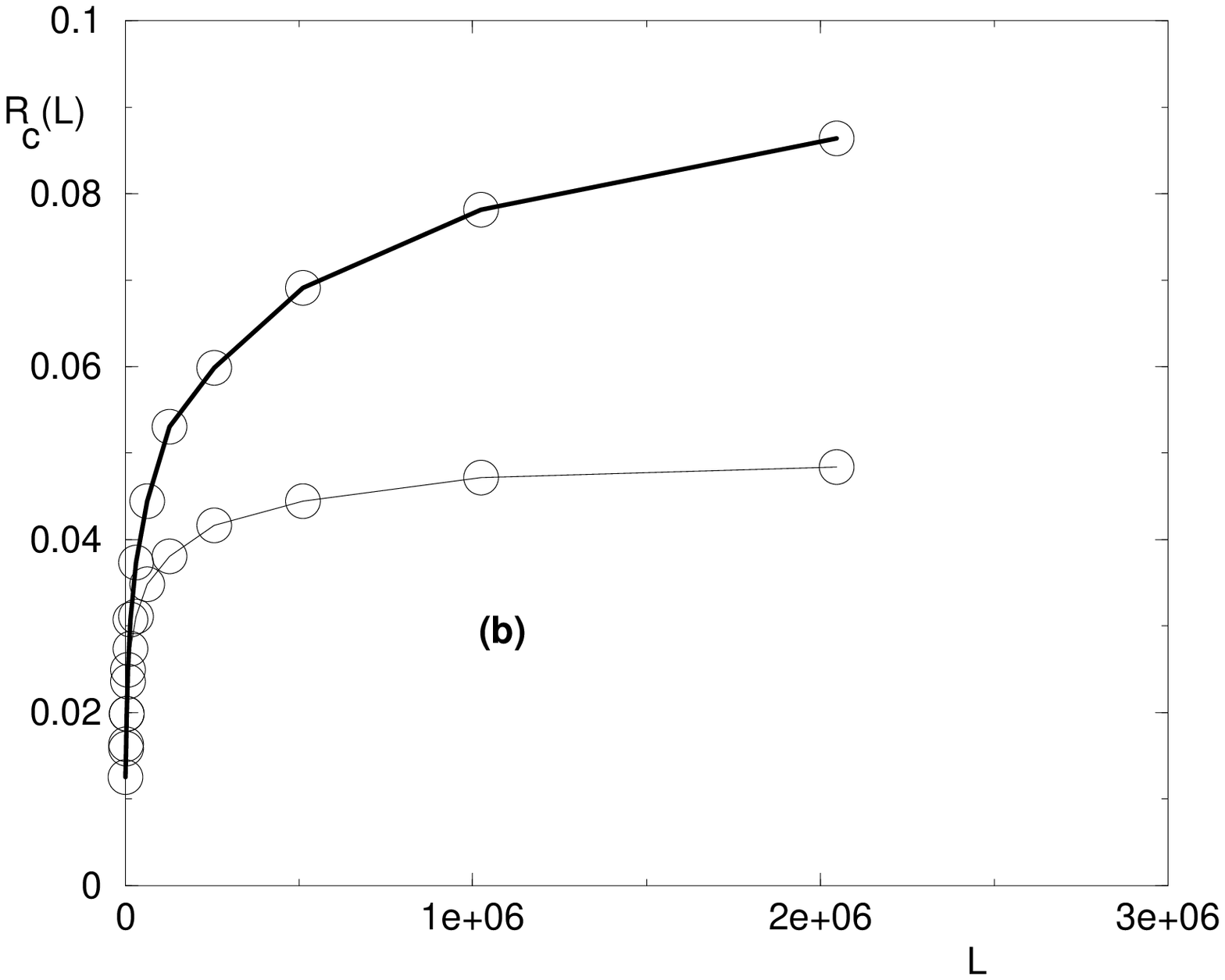}
\caption{ Case c=1.75 : (a)
Ratio $R_{\theta}(T,L)$ for sizes $L/1000=1(\bigcirc),2, 4, 8,
16, 32, 64,128, 256$ (thick line)
(b) The ratios ${\cal R}_c(L)$ of (Eq. \ref{defrcl}), corresponding to
definitions $T_c^{(\theta)}(i,L)$ (thick line) and 
$T_c^{(f)}(i,L)$ (thin line), both grow with $L$.} 
\label{figc175ratio}
\end{figure}

We show on Figure \ref{figc175ratio} (a) the behavior of
the ratio $R_{\theta}(T,L)$ defined in (\ref{ratiotheta}).
It clearly displays a lack of self averaging, as expected at random critical
points (Eq. \ref{nosa}) .

We show on Figure \ref{figc175ratio} (b) the ratios ${\cal R}_c(L)$
of Eq. (\ref{defrcl}) corresponding to the two definitions of
$T_c(i,L)$. They both grow with $L$.

\subsection{ Conclusion on the transition for $c=1.75$}

For relevant disorder $c=1.75$, our conclusion
is that the width $\Delta T_c(L)$ and the shift
 $[T_c(\infty)-T_c^{av}(L)]$ decay with the same new exponent
$L^{-1/\nu_{random}}$ (where $\nu_{random} \sim 2.7 > 2 > \nu_{pure}$)
and there is no self-averaging at criticality, in agreement with the
general predictions (\ref{deltatcrelevant}) and (\ref{nosa}) for
random critical points.

\section{Disorder effects on a first order transition ($c=2.15$)}
\label{c215}

The case $c=2.15$ is of special interest in the context of DNA
denaturation, and the nature of the transition in the presence of
disorder has been under debate recently  \cite{Barbara2,adn2005,Gia_Ton}. 
The numerical studies \cite{Barbara2,adn2005} with bound-unbound
boundary conditions  have found crossings of the energy density
(Fig. 4 of \cite{Barbara2} and Fig. 7b of \cite{adn2005})
and of the contact density (Fig. 6 (b) of \cite{adn2005})
for various sizes $L$.
These results point towards a finite energy density
and a finite contact density at criticality.
However, a recent probabilistic analysis 
\cite{Gia_Ton} of the disordered PS model with $c>2$
has concluded that 
the second derivative of the free-energy 
remains bounded when approaching the critical point
from the localized phase, and
that the order parameter vanishes continuously.
It is thus interesting to reconsider the problem
from the point of view of the distribution of
pseudo-critical temperatures to clarify the nature of the transition.

\subsection{Distribution of pseudo-critical temperatures}

\begin{figure}[htbp]
%\begin{figure}
\includegraphics[height=6cm]{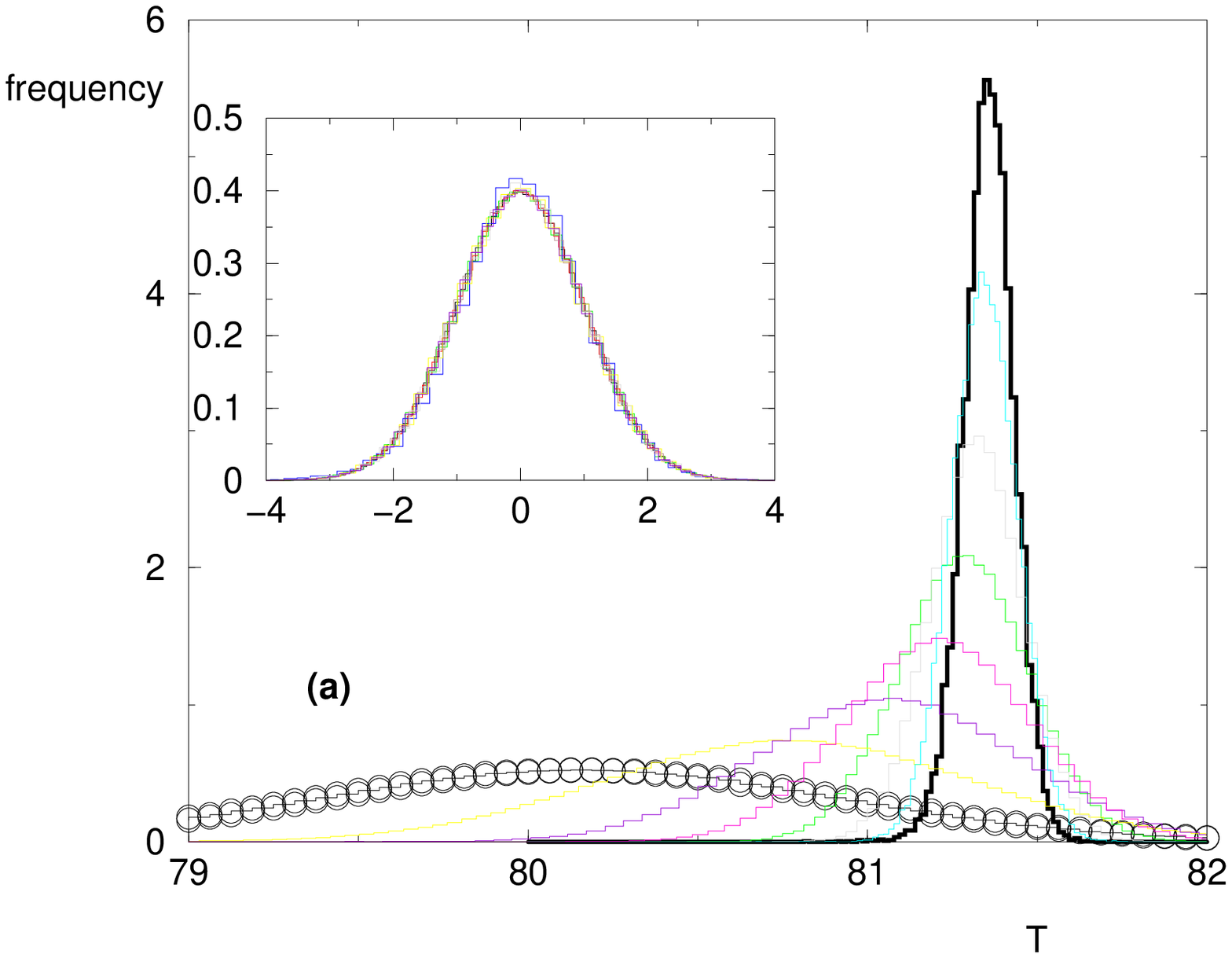}
\hspace{1cm}
\includegraphics[height=6cm]{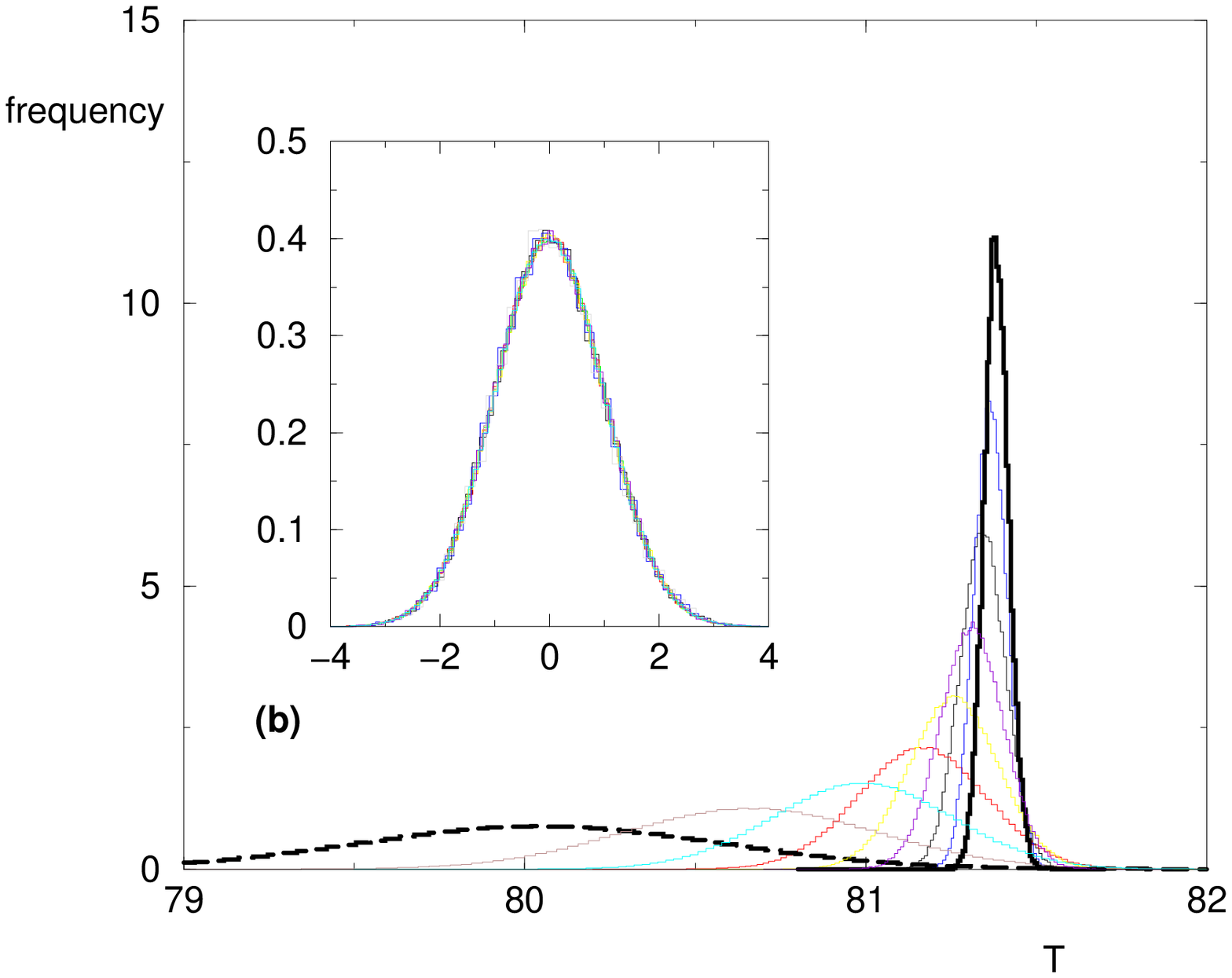}
\caption{ Case c=2.15 :
 Distribution of pseudo-critical temperatures $T_c(i,L)$,
together with the rescaling of Eq. (\ref{rescalinghistotc}) in Inset.
(a) for definition $T_c^{(\theta)}(i,L)$ with samples size
from $L/1000=1 (\bigcirc)$
% ,2, 4, 8, 16, 32, 64, 
to $ 128$ (thick line) 
(b) for definition $T_c^{(f)}(i,L)$ with samples size
from $L/1000= 2$ (dashed line) to $512$ (thick line) } 
\label{figc215histotc}
\end{figure}

The histograms of both definitions of the pseudo critical temperatures
$T_c(i,L)$ are shown on Figure \ref{figc215histotc}(a) and (b). 
Both rescaled distributions shown in the
Insets of Figure \ref{figc215histotc} are Gaussian
(Eq. \ref{rescalinghistotc}).

The averages $T_c^{av}(L)$ behave as
\begin{eqnarray}
T_c^{av(\theta)}(L) && \simeq 81.36- \frac{1254}{L} \nonumber\\
T_c^{av(f)}(L)  && \simeq 81.38- 357 \left(\frac{\ln
L}{L}\right)
\label{shiftc215}
\end{eqnarray}
The two values of $T_c(\infty)$ are close and agree with the value
obtained in Ref. \cite{adn2005} with a different method.
On one hand, according to Eq. \ref{meantc},
 this behavior of the shift seems to indicate
that the exponent is unchanged with respect to
the pure case $\nu_R=1=\nu_P$.  
On the other hand, the width is found to
be much bigger than the shift since

\begin{eqnarray}
\Delta T_c^{(\theta)}(L) && \simeq  23.8 \left(\frac{1}{
L}\right)^{0.49} \nonumber\\
\Delta T_c^{(f)}(L) && \simeq  23.2 \left(\frac{1}{
L}\right)^{0.49}
\label{deltatc215}
\end{eqnarray}

This behavior corresponds to 
the Central Limit estimation $L^{-1/2}$, as in the Harris argument.

The fact that the shift and the width exhibit different exponents
actually means that two diverging correlation lengths
coexist in the presence of disorder
\begin{eqnarray}
\xi_{shift}(T) && \sim \frac{1}{ (T_c-T)} \nonumber  \\
\xi_{var}(T) && \sim \frac{1}{ (T_c-T)^2}
\label{twoxi}
\end{eqnarray}
The presence of two different correlation length exponents 
was already found in other models, in particular in
the random transverse field Ising chain \cite{daniel},
where the exponent $\nu=2$ governs the decay of 
averaged correlation, whereas $\tilde{\nu}=1$ governs
the decay of typical correlations.
In Sec. VII A of Ref. \cite{daniel}, a scaling analysis 
of disorder effects on first-order phase transitions
in dimension $d$ suggests that the presence of
two different correlation length exponents
$\nu=2/d$ and $\tilde{\nu}=1/d$ should be generic 
in these systems : the exponent $\tilde{\nu}=1/d$
is expected to describe the rounding of the transition in a 
typical sample, whereas $\nu=2/d$ describes the rounding of the transition
of the distribution of samples.  
We will discuss this issue in more details in Section \ref{twoexponents},
after the description of our numerical results on non-self-averaging
properties.

\subsection{ Non-self-averaging properties}

\begin{figure}[htbp]
%\begin{figure}
\includegraphics[height=6cm]{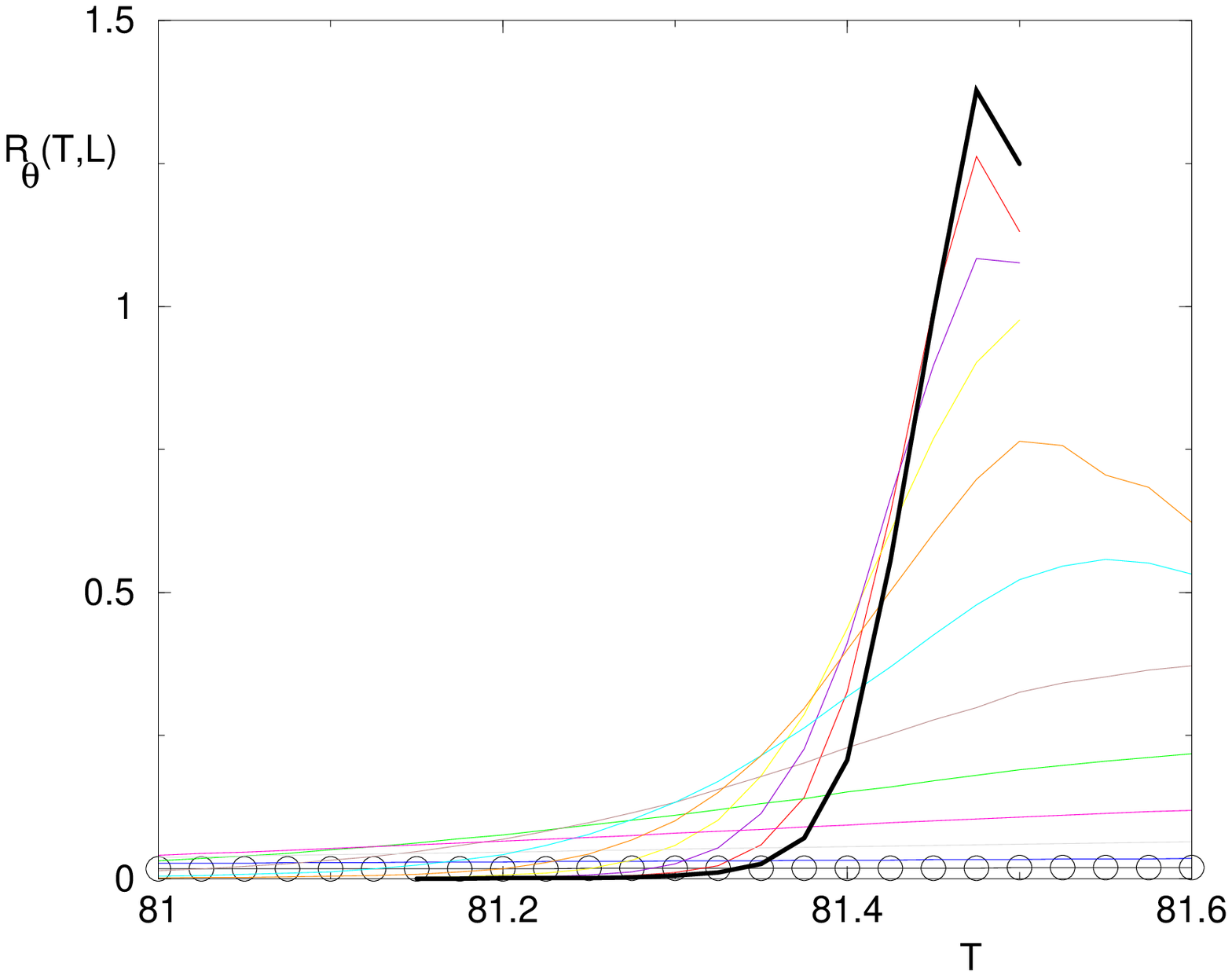}
\hspace{1cm}
\includegraphics[height=6cm]{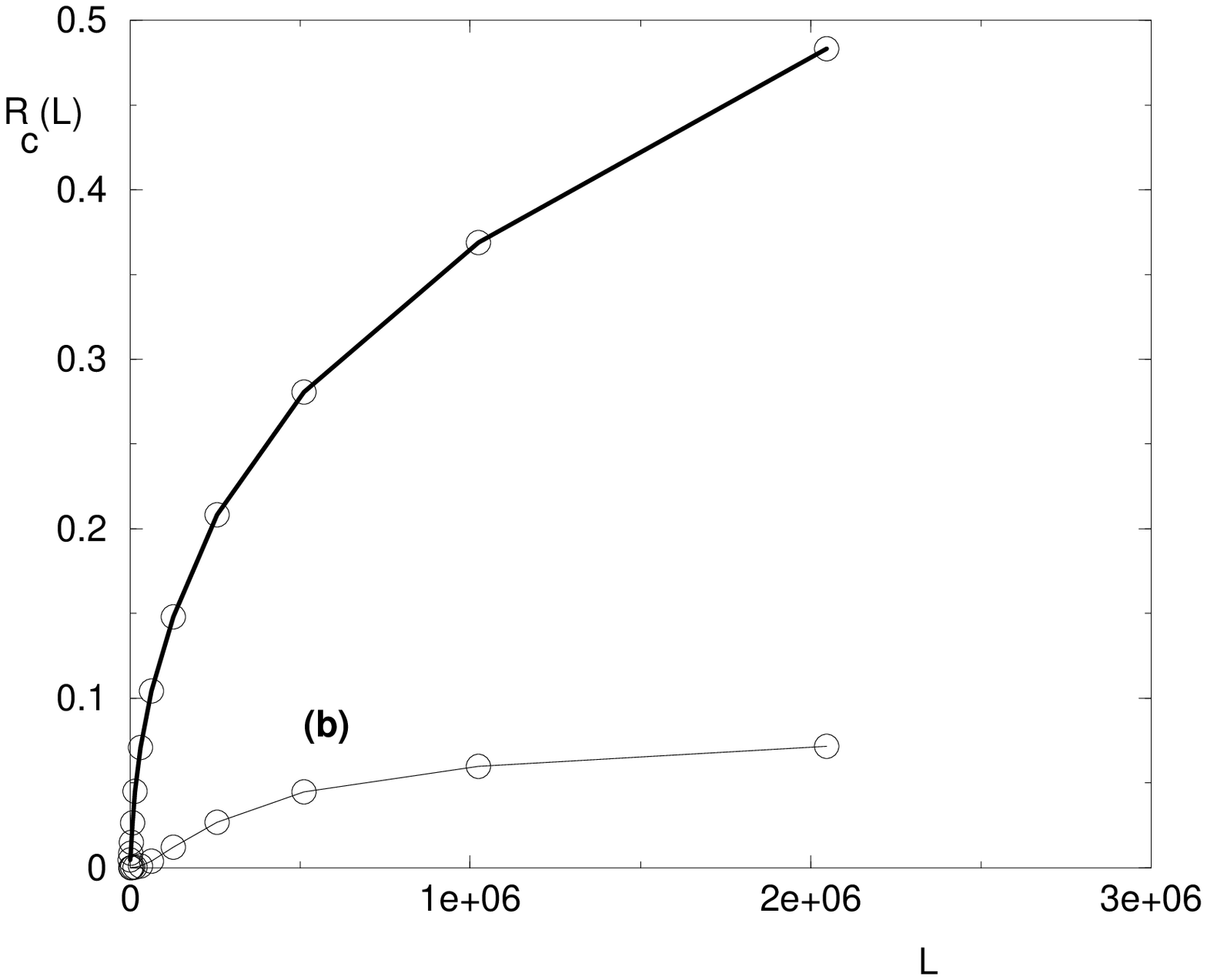}
\caption{ Case c=2.15 :
(a) Ratio $R_{\theta}(T,L)$ for sizes $L/1000=1(\bigcirc)$
% ,2, 4, 8,16, 32, 64,128, 256,512,1024,
to $2048$ (thick line)
(b) The ratios ${\cal R}_c(L)$ of (Eq. \ref{defrcl}), corresponding to
definitions $T_c^{(\theta)}(i,L)$ (thick line) and 
$T_c^{(f)}(i,L)$ (thin line), both grow with $L$.} 
\label{figc215ratio}
\end{figure}

We show on Figure \ref{figc215ratio} (a) the behavior of
the ratio $R_{\theta}(T,L)$ defined in (\ref{ratiotheta}).
It clearly displays a lack of self averaging
at criticality. 

We show on Figure \ref{figc215ratio} (b) the ratios ${\cal R}_c(L)$
of Eq. (\ref{defrcl}) corresponding to the two definitions of
$T_c(i,L)$. They both grow with $L$.

\subsection{ Finiteness of the contact density at criticality}

\label{thetafinite}

\begin{figure}[htbp]
%\begin{figure}
\includegraphics[height=6cm]{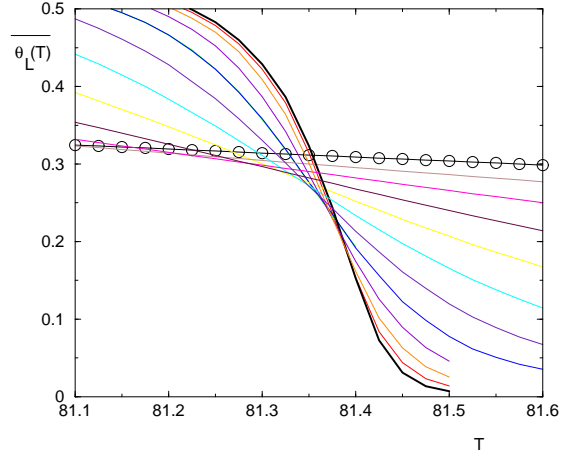}
\caption{ Case c=2.15 :
crossing of the averaged contact density $\overline{\theta_L(T)}$ 
for sizes from  $L/1000=1(\bigcirc)$
% ,2, 4, 8, 16, 32, 64,128, 256,512,1024,
to $2048$ (thick line) .}
\label{figcroisthetamoyc215}
\end{figure}

The dominance of the variance 
(\ref{deltatc215}) over the shift (\ref{shiftc215})
indicate that asymptotically for large $L$, half of the samples $(i,L)$
are still localized at $T_c(\infty)$, whereas the other half
is already delocalized. This suggests that the contact
density is finite at criticality, as we have numerically
found in our previous study \cite{adn2005}.  
However, since Ref. \cite{Gia_Ton} states that the order parameter
vanishes continuously at the transition, we have performed more detailed
calculations.
Figure \ref{figcroisthetamoyc215} (which is more detailed than Figure 6 (b)
of our previous study \cite{adn2005} ) shows the averaged contact
density $\overline { \theta_L(T)}$ as a function of the temperature $T$
for several sizes : the results cross regularly without rescaling.
This points towards a finite contact density at criticality.

 \subsection{ Discussion on the nature of the transition for
 $c=2.15$ in the presence of disorder}

\label{twoexponents}

Our present study shows that, in the presence of disorder,
the transition for $c=2.15$ is an unconventional random critical point
with two different correlation length exponents
$\nu=2$ and $\tilde{\nu}=1$ (\ref{twoxi}).
This is in contrast with usual random critical points, 
arising from second order transitions with
relevant disorder, where the same exponent is expected to govern 
 the width and the shift (Eq. \ref{deltatcrelevant}),
but this is reminiscent of what happens at strong disorder
fixed points \cite{daniel,revue}.
The question is now which correlation exponent appears
in a given observable. In the random transverse field Ising
chain where many exacts results are known for exponents and
scaling distribution functions \cite{daniel},
it is well understood how the two exponents $\nu=2$ and
${\tilde \nu}=1$ govern respectively the averaged/typical correlations.
Here in the disordered PS model, the analog of the correlation
function is the loop distribution. 
To simplify the discussion, let us more specifically
consider the probability of an end-to-end loop of length $L$ in sample (i)
of length $L$, which is directly related to the partition function
$Z^{(i)}_L(T)$ of sample $(i)$
\begin{equation}
P^{(i)}_L(L,T)=\frac {2^L}{L^c}\  \frac {1}{Z^{(i)}_L(T)}
\end{equation}

Introducing for each sample $(i)$ the
difference between the free-energy density 
$F^{(i)}_L(T)/L=-T \ln Z^{(i)}(L,T)/L$
 and the delocalized value $f_{deloc}=-T \ln 2$ 
\begin{equation}
\label{deffi}
f^{(i)}(L,T) \equiv \frac{-T \ln Z^{(i)}(L,T)}{L} +T \ln 2
\end{equation}
 one gets
\begin{equation}
\label{logpl}
{\rm ln }P^{(i)}_L(L,T)=-c \ {\rm ln} L + L \beta f^{(i)}(L,T)
\end{equation}
The self-averaging property of the free energy means
 that $f^{(i)}_L(T)$ converge for large $L$ to a non-random value
$f(T)$ for any sample $(i)$ with probability one
\begin{equation}
 f^{(i)}_L(T)  \operarrow_{L \to \infty}  f(T)
\end{equation}
where $f(T)$ is the free-energy difference between 
the localized phase and the delocalized phase :
$f(T<T_c)<0$ and $f(T>T_c)=0$.
This translates immediately into the corresponding statement (\ref{logpl})
for the logarithm of end-to-end loop probability
\begin{equation}
\frac{ {\rm ln }P^{(i)}_L(L,T) }{L} \operarrow_{L \to \infty} \beta f(T)
\end{equation}
for any sample $(i)$ with probability one.
Since the typical correlation length $\tilde \xi(T)$ 
is usually defined as 
 the decay rate of the logarithm of the correlation,
we obtain here that it is simply given by the inverse of the free-energy 
$f(T)$
\begin{equation}
\label{xitypfree}
 \frac{1}{\tilde \xi(T)} \equiv - \lim_{L \to \infty} 
\left( \frac{ {\rm ln }P^{(i)}_L(L,T) }{L}  \right) = - \beta f(T) 
\end{equation}

Here for $c=2.15$,
we expect from the discussion of the previous Section
\ref{thetafinite} that 
the contact density is finite at criticality : this implies
that the energy is also finite at criticality, and thus
we are led to the conclusion that the free-energy vanishes
linearly
\begin{equation}
\label{freec215}
 f(T)  \operarrow_{T \to T_c^-} (T_c-T)
\end{equation}
The typical correlation length involves the exponent $\tilde \nu=1$
\begin{equation}
\label{resxityp}
\tilde \xi(T)  \operarrow_{T \to T_c^-}
(T_c-T)^{- \tilde \nu} \ \ {\rm with } \ \  \tilde \nu=1
\end{equation}

Let us now consider the decay of the averaged end-to-end loop distribution
that defines an a priori different correlation length $\xi(T)$
\begin{equation}
\label{lnav}
\frac{ {\rm ln } \left( \overline {P^{(i)}_L(L,T) } \right) }{L}
\operarrow_{L \to \infty} - \frac{1}{ \xi(T)}
\end{equation}
This correlation length $\xi(T)$ determines the divergence of
high moments of the averaged loop distribution.
At a given temperature $T<T_c$, these moments will actually
be dominated by the rare samples 
of length $L$ which are already delocalized at $T$, i.e. the 
samples having $T_c(i,L)<T$. Since our numerical results indicate that
the distribution of the pseudo-critical temperature $T_c(i,L)$
is a Gaussian with mean and width given respectively
by Eqs (\ref{shiftc215}) and (\ref{deltatc215}), 
we obtain that the fraction of delocalized samples
presents the following exponential decay in $L$
\begin{equation}
{\rm Prob }[T_c(i,L)<T] \sim  e^{- (T_c^{\infty}-T)^2 L }
\end{equation}
This measure of the rare delocalized samples will govern  
the decay of the averaged
loop distribution, and the correlation length defined in (\ref{lnav})
thus involves the exponent $\nu=2$
\begin{equation}
 \xi(T) \operarrow_{T \to T_c^-} 
(T_c-T)^{ - \nu} \ \ {\rm with } \ \  \nu=2
\end{equation}
in contrast with the typical correlation length (\ref{resxityp}).

To better understand the emergence of two different correlation lengths,
we have numerically measured the distribution over the samples $(i)$
of the free-energy $f^{(i)}(L,T)$ defined in Eq. (\ref{deffi}).
We obtain that for $T<T_c$ 
\begin{equation}
f^{(i)}_L(T) = f(T)+\frac{a_T}{L}+\frac{\sigma_T u_i}{\sqrt L}
\end{equation}
where $a_T$ is temperature dependent and $u_i$ is a Gaussian random
variable of zero mean and of variance $1$ 
\begin{equation}
G(u) = \frac{1}{\sqrt{2 \pi } } e^{- \frac{u^2}{2 }}
\end{equation}
The averaged end-to-end loop distribution then reads (\ref{logpl})
\begin{eqnarray}
\overline {P^{(i)}_L(L,T)}  = \frac{1}{L^c}
  \overline{ e^{ L \beta f^{(i)}(L,T) } }
=  \frac{1}{L^c} e^{ L \beta f(T) }
 \int_{-\infty}^{+\infty} du \ G(u) 
 e^{ {\sqrt L} \beta \sigma_T u }  =  \frac{1}{L^c} e^{ L \beta f(T)
 +L \frac{\beta^2 \sigma^2_T }{2} }
\end{eqnarray}
The difference between
the correlation length $\xi(T)$ (\ref{lnav})
and the typical correlation length (\ref{xitypfree})
is due to the variance $\sigma^2_T$
\begin{equation}
\frac{1}{ \xi(T)} = \frac{1}{\tilde \xi(T)} - \frac{\beta^2 \sigma_T^2}{2} 
\end{equation}
In particular, to obtain the scaling $\frac{1}{ \xi(T)} \sim (T_c-T)^2$
different from $\frac{1}{\tilde \xi(T)} \sim (T_c-T)$, the variance term
in $\sigma_T^2$ has to cancel exactly the leading order
 in $(T_c-T)$ on the left hand-side.

So the picture that emerges of the present analysis is
very reminiscent of what happens at strong disorder fixed points
\cite{daniel,revue} : the exponents $\tilde \nu=1$
and $\nu=2$ govern respectively the decay of
 typical/averaged loop distribution.
Our conclusion is thus that the exponent $\tilde \nu=1$
governs the free-energy (\ref{freec215}) that corresponds to a Lyapunov 
exponent, i.e. it describes the critical behavior of
any typical sample,
whereas the exponent $\nu=2=2/d$ is the finite-size scaling exponent
of Chayes {\it et al} \cite{chayes} and is related to the variance of
the distribution of pseudo-critical temperatures.
Figure 8 of our previous paper \cite{adn2005} may be now interpreted
as follows : for each sample, the critical region has a width
of order $1/L$, whereas the contact density averaged over the samples
decay on a much wider scale $1/\sqrt{L}$ that represents the 
sample-to-sample fluctuations of the pseudo-critical temperatures $T_c(i,L)$.

\section{Summary and discussion}
\label{conclusion}

In this paper, we have used the recent progresses in the 
theory of finite-size scaling in disordered systems
 \cite{domany95,AH,Paz1,domany,AHW,Paz2}
to study the role of disorder in Poland-Scheraga models.
We have obtained that the numerical measure of
the distribution of pseudo-critical temperature $T_c(i,L)$
over the samples is a very powerful tool to elucidate
the true nature of random critical points. The
comparison between the averaged shift $T_c^{av}(L)-T_c(\infty)$ and
the width $\Delta T_c(L)$ clarifies the role of disorder, and allows
one to understand the non self averaging properties of various
observables at $T_c(\infty)$ whenever disorder is relevant.

For $c=1.75$ corresponding to
a second order transition with relevant disorder,
 we have obtained that
both the width $\Delta T_c(L)$ and the shift
$[T_c(\infty)-T_c^{av}(L)]$ decay as
$L^{-1/\nu_{random}}$ with the same new exponent
 $\nu_{random} \sim 2.7 > 2 > \nu_{pure}$.

For $c=1.5$ corresponding to
a second order transition with marginal
disorder, we have obtained that the exponent is unchanged $\nu=2$
with respect to the pure case,
but that disorder is nevertheless relevant :
 the width $\Delta T_c(L) \sim L^{-1/2}$
is of the same order of the shift $T_c^{av}(L)-T_c(\infty)\sim L^{-1/2}$
and this means that thermodynamic observables
(apart from the free-energy) remain distributed at criticality.
We have checked it numerically for the contact density.

For $c=2.15$ corresponding to a first order transition in the pure case,
we have obtained that the width $\Delta T_c(L) \sim L^{-1/2}$
dominates over the shift $T_c^{av}(L)-T_c(\infty)\sim L^{-1}$.
The presence of two correlation length exponents 
 $\nu=2$ and $\tilde \nu=1$ is reminiscent of what happens at
strong disorder fixed points \cite{daniel,revue}, and we have
explained how these two exponents
appear respectively in the typical and averaged loop distributions.

\begin{acknowledgments}

It is a pleasure to thank A. Billoire, G. Giacomin,
J. Houdayer, F. Igl\'oi and J. Jacobsen for discussions. 

\end{acknowledgments}

\end{document}